%% file: arxiv_submission.tex
\documentclass[aps,prb,paper=portrait,twocolumn,superscriptaddress,floatfix,nofootinbib,longbibliography]{revtex4-2}

\usepackage[utf8]{inputenc}
\usepackage{amsmath}
\usepackage{amsfonts}
\usepackage{amssymb}
\usepackage{graphicx}
\usepackage{bbold}
\usepackage{comment}
\usepackage[dvipsnames]{xcolor}
\usepackage{multirow,makecell}
\usepackage{enumitem}
\usepackage{rotating}

\usepackage{bm}
\usepackage{hyperref}
\hypersetup{colorlinks,citecolor=black,linkcolor=blue}
\usepackage[capitalize]{cleveref}

\usepackage{upgreek}

\newcommand{\braket}[1]{{\langle #1\rangle}}
\newcommand{\br}{{\mathbf{r}}}

\begin{document}

\title{Theta electromagnetism in quantum spin ice: Microscopic analysis of improper symmetries}

\author{Gautam K. Naik}
\email{gautamkn@bu.edu}
\affiliation{Department of Physics, Boston University, Boston, Massachusetts 02215, USA}

\author{Jonathan N. Hall\'en}
\email{jonathan\_nilssonhallen@g.harvard.edu}
\affiliation{Department of Physics, Boston University, Boston, Massachusetts 02215, USA}
\affiliation{Department of Physics, Harvard University, Cambridge, Massachusetts 02138, USA}

\author{Chris R. Laumann}
\email{claumann@bu.edu}
\affiliation{Department of Physics, Boston University, Boston, Massachusetts 02215, USA}
\affiliation{Department of Physics, Harvard University, Cambridge, Massachusetts 02138, USA}
\affiliation{Max-Planck-Institut f\"{u}r Physik komplexer Systeme, 01187 Dresden, Germany}

\date{\today}

\begin{abstract}

$U(1)$ gauge theories, including conventional Maxwell electromagnetism, allow $\theta$-terms when parity and time-reversal symmetry are broken.
In condensed matter systems, the physics of $\theta$ as a magnetoelectric response has been explored extensively within the context of topological insulators and multiferroics.
We show how $\theta$-terms can arise in the internal dynamics of the emergent electromagnetism in a $U(1)$ quantum spin liquid.
In its Coulomb phase, the minimal model of pyrochlore quantum spin ice is governed by a six-spin ring exchange Hamiltonian.
We identify the next-order contribution to the microscopic Hamiltonian 
when parity, time-reversal, and all improper spatial symmetries are broken -- a seven-spin term which leads to a two-parameter lattice gauge theory with a $\theta$-electromagnetic phase.
We derive how the seven-spin term is generated perturbatively within each of the three symmetry classes of short-range pyrochlore spin ice.
Within a complete microscopic symmetry analysis, we find that the most general nearest-neighbor Hamiltonians fail to generate the seven-spin term, and one must include next-nearest-neighbor interactions to obtain an emergent $\theta$. 
Using gauge mean-field theory we compute additional contributions to the $\theta$-term from the spinon sector. 
Finally, we determine the conditions required for an internal $\theta$-term to generate a significant external magnetoelectic response.

\end{abstract}

\maketitle

\section{Introduction}
\label{sec:introduction}

Quantum spin ice (QSI) refers to a family of frustrated magnets whose low-energy description features an emergent $U(1)$ gauge theory in its deconfined phase \cite{moessnerThreedimensionalResonatingvalencebondLiquids2003,huseCoulombLiquidDimer2003,hermelePyrochlorePhotons$U1$2004,bentonSeeingLightExperimental2012,savaryQuantumSpinLiquids2017,knolleFieldGuideSpin2019,udagawaSpinIce2021}. 
That is, the QSI phase is governed by the familiar Maxwell action,
\begin{align}
    S_{\text{Maxwell}} = \frac{1}{8\pi \alpha' c'} \int d^3x\, dt\, \mathbf{e}^2 - c'^2 \mathbf{b}^2
    \label{eq:action_maxwell}
\end{align}
but $\mathbf{e}, \mathbf{b}$ are not the usual electromagnetic fields of our universe. 
Rather, they are internal fields related to the long-wavelength magnetization dynamics in the sample. 
Accordingly, the two parameters of Maxwell theory, the speed of `light' $c'$ and coupling $\alpha'$, have nought to do with those of our universe. 
They can even be tuned, as they ultimately depend on the microscopic couplings in the QSI material (or model). 
For example, the dimensionless coupling $\alpha'$ takes values of order $1/10$ in generic QSI models, much larger than the fine structure constant $\alpha \approx 1/137$ of our universe \cite{paceEmergentFineStructure2021}. 
This raises the prospect of QSI materials as rich playgrounds for studying gauge field theories outside the regime of the standard model.

Isotropic Maxwell theory permits an additional coupling associated with the so-called $\theta$-term,
\begin{align}
    S_\theta &= -\frac{\theta'}{4\pi^2} \int d^3x\, dt\, \mathbf{e}\cdot\mathbf{b}
    \label{eq:action_theta}
\end{align}
As a total derivative, $S_\theta$ does not modify the dynamics of `photon' excitations in the bulk of a QSI sample. 
However, it has a range of consequences at the boundary and, in the bulk, it induces the Witten effect, by which magnetic monopoles in $\mathbf{b}$ bind electric monopoles in $\mathbf{e}$ with charge proportional to $\theta'$. 
This is, as far as we know, a `purple cow' effect in our universe, where no fundamental magnetic monopoles have ever been observed. 
In QSI, as a lattice gauge theory built on the lattice of a real magnetic crystal, it is believed that $\mathbf{b}$-monopoles not only exist but are the lowest energy gapped excitations and may be quite important at low temperature \cite{chenDiracsMagneticMonopoles2017,szaboSeeingLightVison2019,laumannHybridDyonsInverted2023}. 
The exotic consequences of $S_\theta$ are thus potentially much more observable.

\begin{figure*}[t]
    \centering
    \includegraphics[width=1\linewidth]{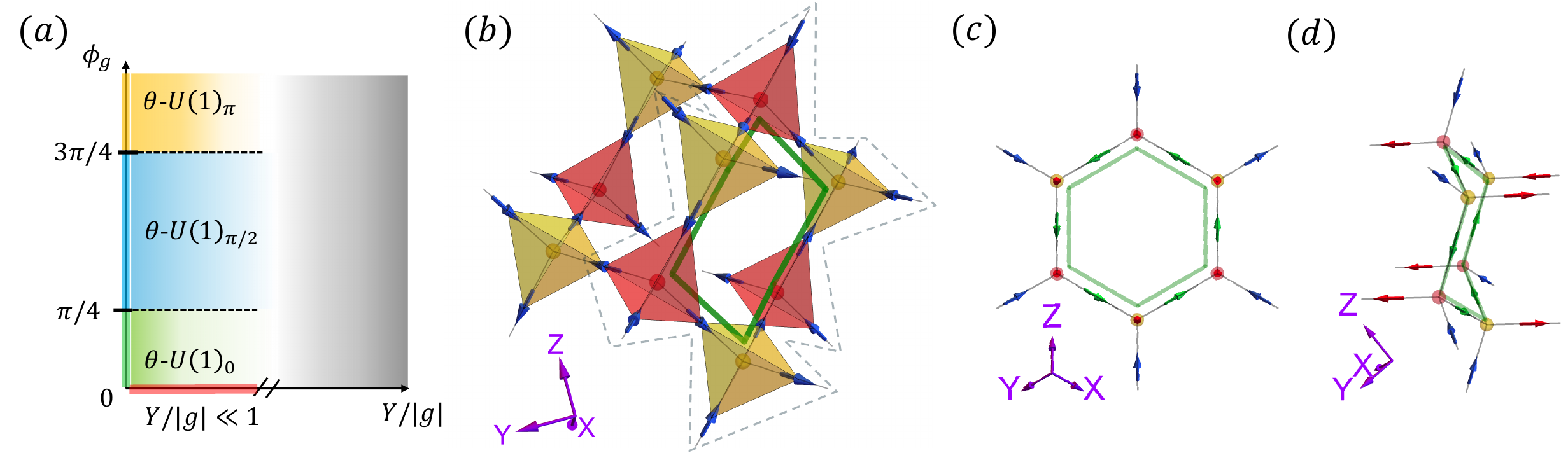}
    \caption{%
     (a) Phase diagram of the extended ring exchange Hamiltonian $H_g+H_Y$ of quantum spin ice in the presence of improper symmetry breaking perturbations, with $g=|g|e^{i\phi_g}$ and real $Y$. 
     The $Y=0$ line has been investigated in previous work considering quantum spin ice on breathing pyrochlore, which has no inversion but still has mirror planes, and consists of four separate deconfined $U(1)$ phases, distinguished by the amount of emergent flux frozen into the ground state in the form of a ``vison crystal'' (indicated by the subscript)~\cite{sandersVisonCrystalQuantum2024}. 
     In this work we show how these deconfined phases gain a $\theta$-term at small $Y/|g|$, focusing on the $\phi_g=0$ line (red highlight). The structure of the model phase diagram at moderate and large values of $Y/|g|$ we leave as an open problem (gray shaded region).
     (b) Pyrochlore spin ice is modeled by a lattice of corner sharing tetrahedra with spins at each vertex subject to strong Ising anisotropy. The centers of the tetrahedra form a diamond lattice with a two-site unit cell, corresponding to upward- and downward-facing tetrahedra (sublattice A/B shown in red/yellow).  
    The system is dominated by a strong $S^z S^z$ interaction, which is minimized by configurations satisfying the ice rule: every tetrahedron has two spins pointing in and two spins pointing out. 
    The most local rearrangement that remains within the ice manifold is a ring exchange, \cref{eq:RingExch}, which flips a loop of head-to-tail spins around the hexagonal plaquettes (green).
    The QSI ground state is a massive superposition of ice rule obeying classical states. 
    (b,c) Different views of the dotted section of (a), with spins along the highlighted hexagonal plaquette $p$ in green. 
    Red spins belong to the set $D(p)$ of spins adjacent and perpendicular to this hexagonal plaquette (see \cref{eq:ham_yterm}); these are parallel to the emergent magnetic field associated with the plaquette. 
    (Color online.)
    }
    \label{fig:pyrochlore_and_ice}
\end{figure*}
In this paper, we consider the question of how $S_\theta$ can be generated microscopically, and how large the dimensionless coupling $\theta'$ is, in the context of the most promising family of QSI candidate materials: the rare-earth pyrochlores (see \cref{fig:pyrochlore_and_ice}). 
On general grounds, we expect $S_\theta$ to be generated so long as symmetry permits it. 
The $\theta$-term is odd under improper space-time transformations, and we must therefore ensure these are broken%
\footnote{Famously, topological insulators generate a $\theta = \pi$ term in the electromagnetic response action despite the presence of time reversal \cite{hasanColloquiumTopologicalInsulators2010,wangTimeReversalSymmetric$U1$2016}. This is permitted because the $\theta$ coupling is periodic in $2\pi$ (for fermionic theories) and thus $\theta \to -\theta$ symmetry permits both $0$ and $\pi$ coupling. We do not consider the topological mechanism for generating non-trivial $S_\theta$ in this manuscript.}.
In the isotropic continuum, this amounts to ensuring that inversion $\mathcal{P}$ and time-reversal $\mathcal{T}$ are broken, but one must be more careful if the underlying space group is crystalline, as in the $F\bar{d}3m$ symmetry of pyrochlore \cite[][App.~B]{burnellMonopoleFluxState2009}. 
As discussed in the long-wavelength setting in \cite{paceDynamicalAxions$U1$2023}, one can break both $\mathcal{T}$ and all improper spatial transformations by considering spin ice on a `breathing pyrochlore' lattice with a coexisting All-In-All-Out (AIAO) antiferromagnetic order. 
Breathing breaks inversion symmetry in the pyrochlore lattice \cite{bentonGroundStateSelection2015,savaryQuantumSpinIce2016,essafiFlatBandsDirac2017,ezawaHigherOrderTopologicalInsulators2018,aoyamaSpinOrderingInduced2019,yanRank2$U1$Spin2020,chernCompetingQuantumSpin2022,sandersVisonCrystalQuantum2024}, but does not remove the mirror planes associated with the achiral tetrahedral group. 
These and time-reversal are broken by the AIAO order \cite{jaubertMonopoleHolesPartially2015,chenMagneticMonopoleCondensation2016,petitObservationMagneticFragmentation2016,lefrancoisFragmentationSpinIce2017,lhotelFragmentationFrustratedMagnets2020}. 
The net effect is to reduce the space group from $F\bar{d}3m$ to $F23$ and permit $S_\theta$. 

Let us now consider the microscopic consequences of the symmetry reduction. 
At energies below the ice energy scale, the minimal model for pyrochlore QSI is given by the ring exchange Hamiltonian \cite{hermelePyrochlorePhotons$U1$2004},
\begin{align}
    H_{g} = -\frac{g}{2} \sum_p S^+_{p_1}S^-_{p_2} S^+_{p_3}S^-_{p_4}S^+_{p_5}S^-_{p_6} + \textrm{h.c.}
    \label{eq:RingExch}
\end{align}
where $p$ enumerates hexagonal plaquettes on the pyrochlore lattice, and the six-spin operator acts on two-level spins pointed head-to-tail around the plaquette $p$ by reversing them.
Ring exchange is the lowest-order off-diagonal process which acts within the manifold of ice states: states which have two spins pointing in and two pointing out of every tetrahedron. 
When the system obeys the space group $F\bar{d}3m$, the Hamiltonian $H_{g}$ is governed by a single real energy scale, $g$, which ultimately sets the speed of light, $c'$, in the Maxwell phase. 
We will review both the derivation of $H_{g}$ from more microscopic spin-spin interactions and why it leads to a deconfined $U(1)$ lattice gauge theory, \cref{eq:action_maxwell}, in \cref{sec:lgt_and_its_continuum,sec:nearest_neighbor_hamiltonian}.

On the breathing pyrochlore lattice, the space group is reduced to $F\bar{4}3m$, and the coupling in \cref{eq:RingExch} is allowed to become complex, $g\to |g| e^{i\phi_g}$. 
The phase diagram of the ring exchange model with a complex $g$ has been explored in recent work \cite{sandersVisonCrystalQuantum2024}: for any $\phi_g$, the ground state of \cref{eq:RingExch} remains a $U(1)$ deconfined phase, but for large enough $\phi_g$ magnetic flux freezes into the ground state to form one of several  ``vison crystals'' (see Fig.~\ref{fig:pyrochlore_and_ice}).
Even with a complex parameter, $H_{g}$ obeys all the spatial symmetries of $F\bar{4}3m$ and $\mathcal{T}$, and it thus cannot generate a $\theta$-term, \cref{eq:action_theta}.
In the remainder of the text, we focus on real $g$ for simplicity but do not expect qualitatively different results in the vison crystal phases.

On reducing the symmetry to $F23$ and breaking $\mathcal{T}$, the next most local off-diagonal term allowed within the ice manifold is a seven-spin term of the form 
(see \cref{fig:pyrochlore_and_ice}),
\begin{align}
    H_{Y} &= i Y \sum_p  S^+_{p_1}S^-_{p_2} S^+_{p_3}S^-_{p_4}S^+_{p_5}S^-_{p_6} \sum_{i \in D(p)} S^z_i + \textrm{h.c.}
    \label{eq:ham_yterm}
\end{align}
where $D(p)$ is the set of pyrochlore lattice sites neighboring the plaquette p on planes directly above and below plane of the plaquette.
$H_Y$ is odd under the improper symmetries of the usual pyrochlore and thus the dimensionless ratio $Y/g$ parametrizes their breaking\footnote{Even though the $F23$ space group allows $Y$ to be complex, we restrict our analysis to a real $Y$. This is because it is the real part that breaks the symmetries to generate an emergent $\theta$-term.}. 
Readers familiar with lattice gauge theory may immediately recognize $H_Y$ as part of a lattice regularization of an $\mathbf{e}\cdot\mathbf{b}$ term. 
By dimensional analysis, we expect that, at least for small $Y$, $H_Y$ generates $S_\theta$ with coupling $\theta' = \frac{Y}{g} f(\frac{Y}{g})$ and $f$ an even, smooth function. 
In \cref{sec:lgt_and_its_continuum}, we derive these connections in more detail in a relaxed rotor approach\footnote{We note that determining the quantum phase diagram of $H_g + H_Y$ as a function of $Y/g$ and the associated coupling $\theta'$ in the deconfined phase in the unrelaxed (2-level spin ice) model would require numerics beyond the scope of this work.}. 
This analysis is valid when the spinon gap is large; that is, the excitations which violate the ice rule are strongly suppressed. 
We discuss the corrections to $\theta'$ due to the gapped spinons and estimate them through a gauge mean-field theory (gMFT) approach in \cref{sec:theta_from_effective_spinon_H}.

The bulk of this paper is devoted to the perturbative derivation of \cref{eq:ham_yterm} from local spin models at the lattice scale for pyrochlore spin ice.
More precisely, there are three classes of pyrochlore magnets, depending on the symmetry of the crystal field ground-state doublet attached to each magnetic atom.
We can label these symmetry classes by $n=1,2,3$ corresponding to whether the doublet behaves like an (1) effective spin-1/2, (2) non-Kramers doublet, or (3) dipolar-octupolar doublet \cite{laumannHybridDyonsInverted2023,rauFrustratedQuantumRareearth2019} (see \cref{sec:symmetry_action_on_effective_spins}). 
It is well-known that only 3 or 4 nearest-neighbor (NN) spin interactions are allowed by full $F\bar{d}3m$ symmetry, depending on the symmetry class $n$. 
The ring exchange Hamiltonian $H_{g}$ can be derived when the NN ice term
\begin{align}
    H_{\textrm{ice}} = J_{zz} \sum_{\langle ij \rangle} S^z_i S^z_j 
    \label{eq:H_ice}
\end{align}
is dominant so that the low energy dynamics can be safely projected into the ice manifold.
In a  Schrieffer-Wolff treatment, the ring exchange coupling $g \sim \frac{J_{NN}^3}{J_{zz}^2}$ is generated at 3rd order, where $J_{NN}$ represents the scale of NN off-diagonal couplings. 

On reducing the symmetry to $F23$ and breaking $\mathcal{T}$, the number of allowed nearest-neighbor terms explodes to $\sim 15$. 
See \cref{tab:symmetry_action} for a complete classification, for each pyrochlore class $n$, organized by behavior under the improper discrete symmetries.
Surprisingly, it turns out that the cornucopia of NN terms are insufficient to generate $H_Y$ perturbatively (see \cref{sec:nearest_neighbor_hamiltonian} for proof). 
One must therefore include next-nearest-neighbor (NNN) two-spin or three-spin terms in order to generate $Y$. 
While we do not exhaustively enumerate these terms 
(of which there are 62), we indicate several representative classes that are sufficient in the lower part of Table~\ref{tab:symmetry_action}. 

The bottom line of the Schrieffer-Wolff analysis is that $Y \sim \frac{J_{NN}^2J_{NNN}}{J_{zz}^2}$ where $J_{NNN}$ is the scale of the off-diagonal next-neighbor or 3-body terms. 
In this approach, $Y$ is formally of the same order as the ring exchange $g$ in the perturbative expansion controlled by the ice scale $J_{zz}$, but whether it can actually be comparable to $g$ depends sensitively on how the background improper symmetry breaking orders couple to the lattice scale physics. That is, how large the relevant symmetry-breaking terms in $J_{NNN}$ are, compared to $J_{NN}$.

As alluded to previously, the $g-Y$ model defined by \cref{eq:RingExch,eq:ham_yterm} can be interpreted as a pure $U(1)$ lattice gauge theory regularized on the pyrochlore lattice. 
The UV regularization of the $\theta$-term has been a topic of research in the lattice gauge theory community for decades~\cite{cardyPhaseStructureZp1982,cardyDualityParameterAbelian1982}. 
Most recently, interesting work has focused on finding UV regularizations which lift various symmetries of the IR $\theta$-electromagnetism explicitly into the UV-completion~\cite{sulejmanpasicAbelianGaugeTheories2019,anosovaPhaseStructureSelfdual2022,paceEmergentHighersymmetryProtected2023}. 
Here, our focus is instead on deriving physical realizations which arise in the phase diagram of pyrochlore magnets.

We emphasize that the $\theta$-term we derive in the context of QSI applies to the \emph{emergent} electromagnetic sector of the theory. 
It exists independently from any notion of external fields. 
This separates the $\theta$-electromagnetism in QSI from other condensed matter systems where $\theta$-terms arise in the magnetoelectric response to applied external fields \cite{nennoAxionPhysicsCondensedmatter2020}. 
Such systems include  multiferroics \cite{riveraShortReviewMagnetoelectric2009}, topological insulators~\cite{sekineAxionElectrodynamicsTopological2021}, Weyl semimetals~\cite{goothAxionicChargedensityWave2019,vazifehElectromagneticResponseWeyl2013,wangChiralAnomalyCharge2013}, and topological superconductors \cite{qiAxionTopologicalField2013}. 
On general symmetry grounds, the presence of $\theta'$ should lead to bulk magneto-electric response as well (see \cref{app:magentoelectric_response_of_QSI}). However, as we will discuss further in \cref{sec:discussion}, the external $\theta$ magnetoelectric response can be parametrically smaller than the internal $\theta'$, as it depends on the extent to which both external electric and magnetic fields couple into the internal magnetic degrees of freedom.

\section{Lattice U(1) gauge theory and its continuum limit}
\label{sec:lgt_and_its_continuum}

In this section we derive the $U(1)$ lattice gauge theory used to study the Coulomb phase of QSI and discuss the form of the $\theta$-term at the lattice scale. 
We also derive the long-wavelength theory and couplings in the relaxed rotor approximation.

The low-energy physics of QSI can be treated by projecting onto the 2-In-2-Out (2I2O) states, which minimize the dominant ice interaction in \cref{eq:H_ice} (see \cref{fig:pyrochlore_and_ice}) \cite{hermelePyrochlorePhotons$U1$2004}.
The simplest QSI Hamiltonian after this projection is one with just the ring exchange term (\cref{eq:RingExch}) that couples the different 2I2O ice rule obeying states.
The ice rule translates to a divergence-free condition on the electric field (Gauss' law) in the $U(1)$ gauge theory description.

To analyze further, it is useful to introduce a rotor relaxation of the spins described by compact $U(1)$ gauge fields $a_i$ and their conjugate momenta $\varepsilon_i$ on the pyrochlore lattice,
\begin{align}
    [a_i,\varepsilon_j]=i \delta_{ij} \,.
\end{align}
We consider rotors with anti-periodic boundary conditions, which constrain ${\varepsilon_i}$ to be half-integer.
The ring exchange Hamiltonian in \cref{eq:RingExch} can be mapped to 
\begin{align}
    H_{\Gamma,\,g} &=\Gamma\sum_i {\varepsilon_i}^2 -g \sum_p \cos (da)_p 
\label{eq:H_emergent_e_and_a_ring}
\end{align}
in the limit $\Gamma\to \infty$, where ${\varepsilon_i}$ is constrained to be $\pm 1/2$ and mapped to the $S^z_i$. Setting $S^\pm_i=\exp\left( \pm i a_i\right)$ maps the ring exchange term in \cref{eq:RingExch} to the cosine of the lattice curl $(da)_p$ in the above expression, which lives on hexagonal plaquettes $p$ of the diamond lattice. These lattice curls are interpreted as the emergent magnetic fields on the lattice, $b_p=(da)_p$. 

In the limit $\Gamma \to 0$, the cosine can be expanded to the quadratic order to obtain the Hamiltonian:
\begin{align}
    H_{\Gamma \ll g} &=\Gamma\sum_i {\varepsilon_i}^2 +g \sum_p b_p^2 
\label{eq:H_emergent_e_and_a_small_gamma}
\end{align}
This has the form of the Hamiltonian for Maxwell electromagnetism on a lattice (see \cref{sec:Lagrangian_with_theta_term}). It can be used to obtain the scaling of the emergent fine structure constant, ${\alpha'} \propto \sqrt{\Gamma/g}$, and the emergent speed of light, ${c'}\propto \sqrt{\Gamma g}\, l$ of the $U(1)$ gauge theory (where $l$ is the lattice spacing). 
This will be explicitly shown in \cref{sec:continuum_limit_for_gamma_zero}.
However, this scaling no longer applies in the limit $\Gamma\to\infty$, where the only relevant energy scale is $g$, which leads to the emergent fine structure constant asymptotically approaching the value ${\alpha'} \approx 0.1$ \cite{paceEmergentFineStructure2021} and the emergent speed of light scaling as ${c'}\propto g l$ \cite{shannonQuantumIceQuantum2012,kwasigrochSemiclassicalApproachQuantum2017,paceEmergentFineStructure2021}.

Having established how the emergent electromagnetic description arises in QSI we will now demonstrate how the term in \cref{eq:ham_yterm} leads to a $\theta$-term coupling between ${\varepsilon_i}$ and $b_p$, and estimate the scaling of ${\theta'}$.

\subsection{$\theta$-term in the lattice gauge theory}

As discussed in \cref{sec:introduction}, the ring exchange term in \cref{eq:RingExch} is the lowest-order off-diagonal term within the ice manifold and, with appropriate symmetry reduction, the seven-spin term in \cref{eq:ham_yterm} is the lowest-order term odd under the improper symmetries. When expressed in terms of the emergent $U(1)$ lattice gauge fields, this seven-spin term takes the form:
\begin{align}
  H_{Y} &= 2 Y \sum_p\sum_{i\in D(p)} {\varepsilon_i}\sin (da)_p \, .
\label{eq:H_Y_emergent_e_and_a}
\end{align}
Here, and in \cref{eq:ham_yterm}, $D(p)$ denotes the set of six edges that have endpoints on the boundary of the hexagonal plaquette $p$, each oriented perpendicular to the plane of the plaquette (see \cref{fig:pyrochlore_and_ice}).
These specific edges are considered since they are the ones with emergent electric fields parallel to the emergent magnetic field of the plaquette; this allows us to study the isotropic $\theta$-response of the system.
Other seven-spin terms -- with $S^z$ on other pyrochlore sites neighboring the six-spin ring exchange -- result in an anisotropic magnetoelectric response and are allowed only when the rotational symmetries of the tetrahedral point group $T$ are broken. These will not be discussed further.

With the Hamiltonian, $H=H_{\Gamma,\,g}+H_{Y}$ (in \cref{eq:H_emergent_e_and_a_ring,eq:H_Y_emergent_e_and_a}), we can use Hamilton's equations to obtain
\begin{align}
    \dot{a}_i= \frac{\partial H}{\partial {\varepsilon_i}}&= 2\Gamma {\varepsilon_i}+2Y \sum_{p\in D(i)}\sin(da)_p ,
    \label{eq:a_dot_lgt}
\end{align}
and calculate the Lagrangian of the system:
\begin{align}
    L&=\frac{1}{4\Gamma} \sum_i \dot{a}_i^2 - \frac{g}{2} \sum_p\cos(da)_p -\frac{Y}{\Gamma} \sum_p \sum_{i\in D(p)} \dot{a}_i \sin(da)_p  \nonumber\\
    &\quad +\frac{Y^2}{\Gamma} \sum_i \sum_{p, p'\in D(i)} \sin(da)_p \cdot \sin(da)_{p'} \, ,
\label{eq:L_emergent_e_a}
\end{align}
where $D(i)$ is the set of all the plaquettes that share a lattice point with the edge $i$, each in the plane perpendicular to the edge.

Note that the emergent electric field, $\mathbf{e}$, is the negative time derivative of $\mathbf{a}$ (in the Coulomb gauge). 
From \cref{eq:a_dot_lgt}, we can see that when $Y\neq0$, $\varepsilon$ represents a combination of electric and magnetic fields rather than purely an electric field. Furthermore, the ice rule for the spins in QSI translates to Gauss's law for the canonical momentum $\varepsilon$, not the electric field.

For the considered Hamiltonian, in the limit that it maps to the spin system, i.e. $\Gamma\to \infty$, the only relevant energy scale is $g$, and the only dimensionless parameter is $Y/g$. 
Because $Y$ is odd under both $\mathcal{T}$ and improper spatial transformations, it is guaranteed that 
\begin{align}
    {\theta'} = \frac{Y}{g} f\left(\frac{Y}{g}\right) \, ,
    \label{eq:theta_scaling_large_Gamma}
\end{align}
where $f(x)$ is a smooth scaling function that is even. 

Note that the discussion in this section assumes that the electric ($\varepsilon$-) charges of the lattice gauge theory are sufficiently gapped, and that we are only interested in the effective theory at energies that are smaller than these gaps. 
In particular, we have projected out all the electric charges when projecting onto the ice manifold.
If one allows electric charges, which have an energy gap $\Delta=J_{zz}/2$, ${\theta'}$ receives sub-leading corrections proportional to $Y/\Delta$. 
We discuss this sub-leading contribution and a gauge mean-field theory (gMFT) approach to estimate it in \cref{sec:theta_from_effective_spinon_H}.

\subsection{Continuum limit for $\Gamma\to 0$}
\label{sec:continuum_limit_for_gamma_zero}

The limit $\Gamma \to 0$ permits a naive continuum limit to be used to derive the $\theta$-Maxwell theory in the IR. 
The low energy states are those with large fluctuations in $\varepsilon$ and small fluctuations of $a$ around the minimum of the cosine, i.e. $a_i=0$. 
By expanding around the state with $a_i=0$, we obtain
\begin{align}
    H&=\Gamma\sum_i {\varepsilon_i}^2 + \frac{g}{2} \sum_p  (da)^2 + 2 Y \sum_p\sum_{i\in D(p)} {\varepsilon_i} (da)_p 
    \, .
    \label{eq:H_emergent_e_a_quadratic}
\end{align}
In the same limit, the Lagrangian expanded to quadratic order is
\begin{align}
    &L=\frac{1}{4\Gamma} \sum_i \dot{a}_i^2 - \left( \frac{g}{2}-6 \frac{Y^2}{\Gamma} \right) \sum_p(da)_p^2 \nonumber\\
    & -\frac{Y}{\Gamma} \sum_p \sum_{i\in D(p)} \dot{a}_i \cdot (da)_p +\frac{Y^2}{\Gamma} \sum_i \sum_{p\neq p'\in D(i)} (da)_p \cdot (da)_{p'}
    \, .
\label{eq:L_emergent_e_a_quadratic}
\end{align}

This quadratic theory admits a naive continuum limit.
We define a continuous vector potential field ${\boldsymbol{\alpha}}({\mathbf{x}})$ by:
\begin{align}
    a_i=\int_i {\boldsymbol{\alpha}} \cdot d{\mathbf{x}} \approx l \ \hat{u}_i \cdot {\boldsymbol{\alpha}}(\bar{{\mathbf{r}}}_i)
    \, ,
\end{align}
where $l$ is the distance between two diamond lattice sites, $\hat{u}_i$ is the unit vector along edge $i$ pointing out of a point in sublattice $A$ of the diamond lattice, and $\bar{{\mathbf{r}}}_i$ is the midpoint of the edge. 
Working to leading order in $l$, we can further express the lattice curl as
\begin{align}
    (da)_p \approx \mathcal{A}_h\  \hat{u}_p \cdot \boldsymbol{\nabla}\times {\boldsymbol{\alpha}}(\bar{{\mathbf{r}}}_p) 
    \, ,
\end{align}
where $\mathcal{A}_h=\frac{3\sqrt{ 3 }}{2}l^2$ is the area of the hexagonal plaquette projected onto the plane perpendicular to $\hat{u}_p$ and $\bar{{\mathbf{r}}}_p$ is the center of the plaquette. 
With the above substitutions, the  lowest-order terms of the Lagrangian in the continuum limit are (see \cref{sec:discrete_sums_to_continuum_integrals}):
\begin{align}
    L&=\int \mathrm{d}^3x \left( \frac{1}{\Gamma} \frac{1}{3\sqrt{ 2 }l} \dot{\boldsymbol{\alpha}}^2 -\left( \frac{g}{2} -  \frac{36Y^2}{\Gamma} \right) \frac{9\, l}{4\sqrt{ 2 }} ({\boldsymbol{\nabla}}\times \boldsymbol{\alpha})^2   \right. \nonumber\\
    &\hskip0.5\linewidth  \left.- \frac{Y}{\Gamma} \frac{3\sqrt{ 3 }}{\sqrt{2}} \;  \dot{ {\boldsymbol{\alpha}}} \cdot {\boldsymbol{\nabla}}\times {\boldsymbol{\alpha}} \right) 
\label{eq:L_continuum}
\end{align}
From the above Lagrangian, we extract the emergent couplings (see \cref{eq:L_free_EM}):
\begin{align}
    {\alpha'} &= \frac{4}{\sqrt{3}} \sqrt{\frac{\Gamma}{g- \frac{72 Y^2}{\Gamma}}}\\
    {c'} &= \frac{3\sqrt{3}}{2\sqrt{2}} \sqrt{\Gamma \left(g-\frac{72Y^2}{\Gamma} \right)} \ l\\
    {\theta'} &=6\sqrt{6}\pi^2\frac{Y}{\Gamma}
\label{eq:theta_in_Y_Gamma}
\end{align}
We remind the readers that the above expression gives the value of ${\alpha'},\ {c'}$ and ${\theta'}$ for $\Gamma \ll g$, and that the mapping of the lattice gauge theory to QSI holds in the limit $\Gamma\to \infty$.

\begin{widetext}

\section{Generating $Y$: Pyrochlore symmetry analysis}
\label{sec:theta_in_qsi}

In this section, we describe the ingredients required in a QSI Hamiltonian for the system to have a non-zero ${\theta'}$ and compute the $Y$-coefficient perturbatively. We first write out the most general nearest-neighbor (NN) pyrochlore QSI Hamiltonian with broken inversion ($\mathcal{I}$), time-reversal ($\mathcal{T}$) and improper tetrahedral point group ($\mathcal{M}$) symmetries. 
With a symmetry analysis of the perturbative expansion of the Hamiltonian, we show that the NN terms are insufficient to generate a $Y$-term in \cref{sec:nearest_neighbor_hamiltonian}.
We then introduce next-nearest-neighbor (NNN) three-spin and two-spin terms capable of perturbatively generating the $Y$-term and compute the $Y$-coefficient of the system in \cref{sec:three-spin_terms_generate_theta,sec:long_range_two_spin_terms_that_generate_theta}.

In most of the main text and the Appendix, we consider pyrochlore QSI with dipolar-octupolar doublets ($n=3$), chosen for their simple transformation properties under the action of the rotational symmetries of the tetrahedral point group (see \cref{eqs:doublet_spin_symmetry_transformations}). 
Similar analysis can be done for the effective spin 1/2 doublets ($n=1$) and the non-Kramers doublets ($n=2$), and the results for these classes of doublets are collected in \cref{sec:effective_spin_half_and_non_kramers_doublets}.

\subsection{Nearest-neighbor Hamiltonian}
\label{sec:nearest_neighbor_hamiltonian}

The symmetries of QSI on a pyrochlore lattice can include the achiral tetrahedral point group ($T_d$) about a diamond site, inversion ($\mathcal{I}$) about a pyrochlore lattice site, time reversal ($\mathcal{T}$) and translational symmetry.
With all of these symmetries present, the dipolar-octupolar doublet QSI Hamiltonian with NN couplings has 4 real parameters:
\begin{align}
H_{T_d}&=\sum_{\braket{ ij}}\left\{ J_{zz} S^z_i S^z_j - J_\pm(S^+_i S^-_j + S^-_i S^+_j) + J_{++}(S^+_i S^+_j+ S^-_i S^-_j) + J_{z+} \left[  (S^z_i S^+_j + S^+_i S^z_j)+(S^z_i S^-_j + S^-_i S^z_j) \right]   \right\} 
\label{eq:H_class3_S4}
\end{align}
When the off-diagonal interactions are weak compared to the Ising term (i.e., $J_{\pm},\ J_{++},\ J_{z+} << J_{zz}$) the low-energy physics of this Hamiltonian is described by dynamically connected ice rule obeying spin configurations. 
The minimal model of this quantum dynamics is the ring exchange Hamiltonian in \cref{eq:RingExch}. 
The six-spin ring exchange term appears at third order in perturbation theory ($g=24J_{\pm}^3/J_{zz}^2$) and is the minimal off-diagonal term that connects ice states \cite{hermelePyrochlorePhotons$U1$2004}. For small $J_{\pm}$, the system is predicted to enter the Coulomb quantum spin liquid phase \cite{savaryCoulombicQuantumLiquids2012,haoBosonicManybodyTheory2014}.

To see a non-zero $\theta'$ within the Coulomb quantum spin liquid phase, the Hamiltonian is required to break all the symmetries under which $\theta'$ is odd. These include $\mathcal{I}$, $\mathcal{T}$ and all the mirror symmetries $\mathcal{M}$ (the improper elements of the $T_d$ point group symmetry). 
Restricting our Hamiltonian to have single spin and NN two-spin terms with just the tetrahedral point group ($T$, the proper subgroup of $T_d$) and translational symmetry leads to 15 free parameters:
\begin{align}
H_0=\sum_{\sigma\in\{A,B\}}\sum_{\braket{ ij}_\sigma} \bigg\{& J_{zz,\sigma}\, S^z_i S^z_j - J_{\pm,\sigma}(S^+_i S^-_j + S^-_i S^+_j) + \left( \left(J_{++,\sigma} + i\, J_{--,\sigma}\right) S^+_i S^+_j + \textrm{h.c.} \right)  \nonumber \\ 
& + \left(\left(J_{z+,\sigma} + i\, J_{z-,\sigma} \right)  (S^z_i S^+_j + S^+_i S^z_j) + \textrm{h.c.} \right)   \bigg\}
+\sum_{i} \bigg\{h_z\ S^z_i +\frac{h_x}{2}\ (S^+_i+S^-_i) -i \frac{ h_y}{2}\ (S^+_i-S^-_i) \bigg\}
\label{eq:H_class3_2spin_A4}
\end{align}
Here, $\langle ij\rangle_A$ ($\langle ij\rangle_B$) refer to the set of edges on the upward (downward) facing tetrahedra of the pyrochlore (see \cref{fig:pyrochlore_and_ice}). 
We reparameterize the above couplings by introducing breathing parameters $\kappa_s$ to symmetry-diagonalize the action of inversion:
\begin{align}
&J_{s,\sigma} = J_s(1+\eta_{\sigma}\kappa_s) 
\label{eq:J_kappa_reparameterization}\\
&\text{where }\eta_{A}= 1 \text{ and } \eta_B=-1.
\end{align}
Here, $s\in\{zz,\, \pm,\, ++,\, --,\, z+,\, z-\}$ -- i.e., $J_s$ in the above expression can be any of the six different types of NN couplings. The action of $\mathcal{I},\mathcal{T}$ and $\mathcal{M}$ on the symmetry-diagonalized parameters are summarized in the first group of rows of \cref{tab:symmetry_action}.

The table also shows action of modified mirror symmetries, $\mathcal{\tilde{\mathcal{M}}}$, which we define by the following action for dipolar-octupolar doublet spins (see \cref{sec:symmetry_action_on_effective_spins} for the general definition):
\begin{align}
    \tilde{\mathcal{M}}_R: & S^z_r \rightarrow S^z_{R^{-1}r} \nonumber\\
    & S^\pm_r \rightarrow  S^{\pm}_{R^{-1}r} 
    \label{eq:define_modified_mirror_symmetry_n_3}
\end{align}
where $R$ is a improper rotation of the $T_d$ point group. 
The $Y$-term in \cref{eq:ham_yterm} is odd under the action of $\mathcal{\tilde{\mathcal{M}}}$. Any nearest-neighbor Hamiltonian we write which satisfies the $F23$ space group symmetry is, however, even under $\mathcal{\tilde{\mathcal{M}}}$. In other words, no nearest-neighbor Hamiltonian will perturbatively generate $H_Y$, as it cannot break this ``accidental'' symmetry. We shall have to consider further-neighbor interactions in order to break $\mathcal{\tilde{\mathcal{M}}}$ and generate an emergent $\theta'$.

\subsection{Three-spin terms that generate $\theta$}
\label{sec:three-spin_terms_generate_theta}

\begin{figure}
    \centering
    \includegraphics[width=0.9\linewidth]{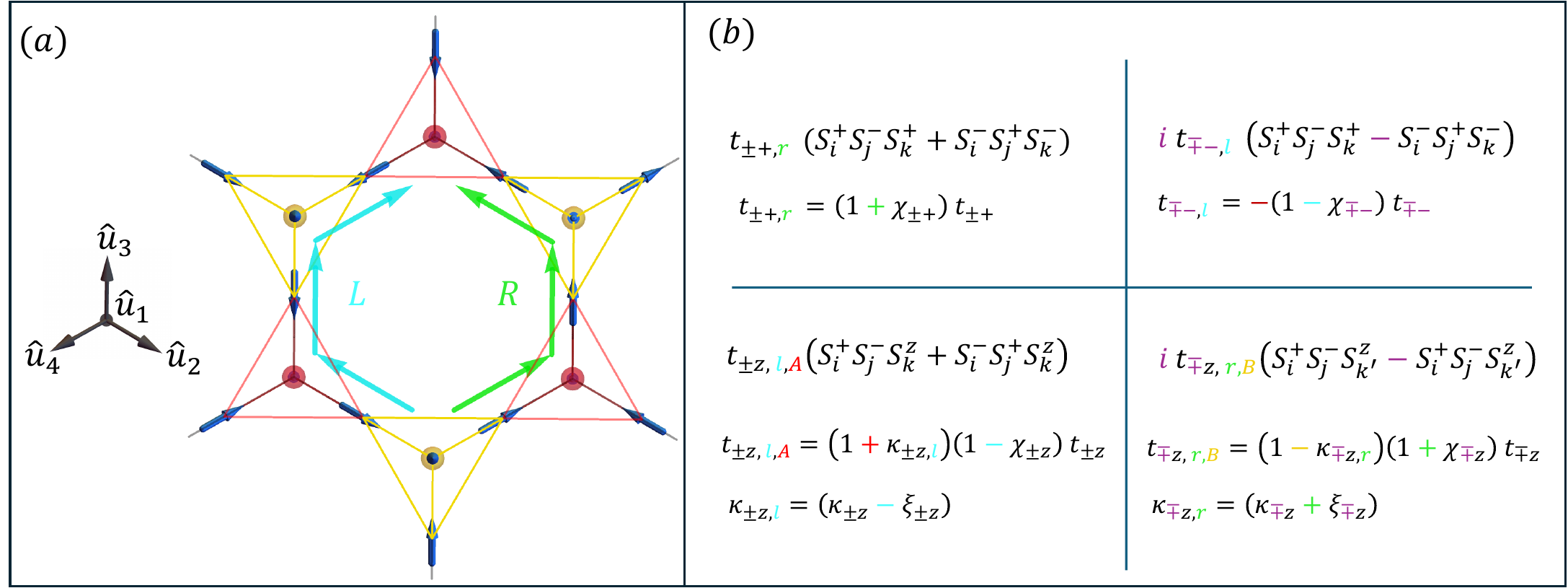}
    \caption{Next-nearest-neighbor (NNN) three-spin terms on the pyrochlore lattice. Triplets $\braket{ijk}$ that connect NNN sites $i$ and $k$ lie on hexagonal plaquettes of the pyrochlore lattice. All NNN triplets $\braket{ijk}$ can be separated into four groups distinguished by two aspects: 
    (i) the edge $\braket{ij}$ belongs to an upward facing tetrahedron (red), $\braket{ijk}_A$, or a downward facing tetrahedron (yellow), $\braket{ijk}_B$, of the pyrochlore, and
    (ii) they are left (cyan), $\braket{ijk'}_{\sigma,L}$, or right (green),$\braket{ijk}_{\sigma,R}$, oriented. 
    The orientation is right (left) if the sign of the scalar triple product $(\hat{u}_i\times\hat{u}_j)\cdot\hat{u}_k$ is positive (negative), where $\hat{u}_i$ is the direction of the easy axis pointing out of the upward facing A tetrahedron at site $i$. (a) A hexagonal plaquette of the pyrochlore with a choice of left and right-oriented NNN triplets shown. (b) A few of the Hamiltonian parameter definitions with colors highlighting the effect of each of the four groups (refer \cref{eq:define_kappa_xi,eq:define_chi}). The purple color represents that the parameter is the imaginary part of a parameter (its subscripts are the sign-flipped counterparts of the real part, see \cref{eq:define_bold_complex_parameters}). The negative sign in brown (in the upper left quadrant) is used to indicate that it is only present when the parameter is both left oriented and an imaginary part (see \cref{eq:define_chi}).
    $\kappa$ captures the sublattice dependence of  $t$-parameters, $\chi$ captures the orientation dependence of $t$-parameters and $\xi$ captures the orientation dependence of the $\kappa$-parameters.  
    (Color online.)   }
    \label{fig:nnn_rl_parameter_description}
\end{figure}

Since NN terms in the Hamiltonian are insufficient to generate the $Y$-term in \cref{eq:ham_yterm}, we consider next-nearest-neighbor (NNN) two-spin terms and three-spin terms. Although characterizing NNN three-spin interactions in materials might be more challenging than NNN two-spin interactions, we choose to analyze them first as they permit simpler perturbative calculations. 

While several possible terms can be added, we only consider those that help generate the term in \cref{eq:ham_yterm}.
Breaking $\mathcal{T}$ and $\mathcal{I}$, but retaining full $T_d$ point group symmetry, we can add the terms:
\begin{align}
H_{3,T_d}=&  \sum_{\sigma\in\{A,B\}}\sum\limits_{\braket{ijk}_\sigma}
\left(t_{\pm z,\sigma}+i\,t_{\mp z,\sigma}\right) S_i^+ S_j^- S_{k}^z +
\left(t_{\pm +} +i\,t_{\mp -} \right) \sum\limits_{\braket{ijk}_A} S_i^+ S_j^- S_{k}^++ \textrm{h.c.}\, ,
\label{eq:H3_class3_S4}
\end{align}
where $\braket{ijk}_\sigma$ represents the set of all NNN triplets, with edge $\braket{ij}\in\braket{ij}_\sigma$, on the pyrochlore lattice. 
When we further break the $T_d$ point group symmetry down to $T$ (to break $\mathcal{M}$ symmetries) the terms in the above Hamiltonian get split into left and right-oriented three-site interactions, and the resulting Hamiltonian is 
\begin{align}
H_3= 
& \sum_{\sigma\in\{A,B\}} 
\sum_{\braket{ijk}_{\sigma,R} }\left(t_{\pm z,r,\sigma} + i\, t_{\mp z,r,\sigma} \right)  S_i^+ S_j^- S_{k}^z 
- \sum_{\sigma\in\{A,B\}}\sum_{\braket{ijk'}_{\sigma,L} }\left(t_{\pm z,l,\sigma} + i\, t_{\mp z,l,\sigma} \right)    S_i^+ S_j^- S_{k'}^z  
\nonumber \\
&+  \left(t_{\pm +,r}+ i\, t_{\mp -,r}\right)  \sum_{\braket{ijk}_{A,R}}   S_i^+ S_j^- S_{k}^+ 
- \left( t_{\pm+,l}+i\, t_{\mp -,l}\right)  \sum_{\braket{ijk'}_{A,L}}  S_i^+ S_j^- S_{k'}^+  
+ \textrm{h.c.} 
\label{eq:H3_class3_A4}
\end{align}
where, $\braket{ijk'}_{\sigma,L}$ and $\braket{ijk}_{\sigma,R}$ represent the left and right oriented NNN triplets, respectively, with edge $\braket{ij}\in\braket{ij}_\sigma$ (see \cref{fig:nnn_rl_parameter_description}). 
The above Hamiltonian has 12 parameters, which transform non-trivially under the action of $\mathcal{I},\mathcal{T}$ and $\mathcal{M}$. To separate the action of $\mathcal{I}$, we define:
\begin{align}
t_{s,r,\sigma}&=(1+\eta_{\sigma}\ \kappa_{s,r})\ t_{s,r} &\text{ and } &&\kappa_{s,r}&=\kappa_s+\xi_s \nonumber \\ 
t_{s,l,\sigma}&=(1+\eta_{\sigma}\ \kappa_{s,l})\ t_{s,l}  &&&\kappa_{s,l}&=\kappa_s-\xi_s 
\label{eq:define_kappa_xi}
\end{align}
where, $s\in \{(\pm z),(\mp z) \}$. With such a reparametrization, the action of $\mathcal{I}$ is completely characterized by $\kappa_s$ while the action of $\mathcal{M}$ on $\kappa_{s,r}$ and $\kappa_{s,l}$ is characterized by $\xi_s$. 
To separate the action of $\mathcal{M}$ on the remaining left and right oriented parameters, we define $\chi_s$ by:
\begin{align}
t_{s,r}&=(1+\chi_s)\ t_s  &\text{ and } &&t_{-s,r}&=(1+\chi_{-s})\ t_{-s} \nonumber \\ 
t_{s,l}&=(1-\chi_s)\ t_s  &&&t_{-s,l}&=(-1+\chi_{-s})\ t_{-s} 
\label{eq:define_chi}
\end{align}
where, $s\in \{(\pm z), (\pm+) \}$ and $-s$ is the counterpart of $s$ with all its signs flipped. 
We further define
\begin{align}
\mathbf{t}_{s}= t_{s} + i \ t_{-s} 
\label{eq:define_bold_complex_parameters}
\end{align}
where, $s\in \{(\pm z),(\pm z,r),(\pm z,l), (\pm+),(\pm+,r),(\pm+,l) \}$ for convenience. 
The action of $\mathcal{I}$, $\mathcal{T}$, $\mathcal{M}$ and $\tilde{\mathcal{M}}$ symmetries on the terms in the Hamiltonian is shown in \cref{tab:symmetry_action} and a schematic description of the various parameter definitions is shown in \cref{fig:nnn_rl_parameter_description}.


\begin{table*}
\begin{tabular}{|c|c|c|c|c|c|c|c|c|c|c|c|c|c|c|c|c|c|c|c|}
\hline
\rule{0pt}{11pt}&\multicolumn{1}{|c|}{Effective spin 1/2 $(n=1)$} &$\mathcal{I}$&$\mathcal{T}$&$\mathcal{M}$&$\tilde{\mathcal{M}}$
&&\multicolumn{1}{|c|}{Non-Kramers $(n=2)$} &$\mathcal{I}$&$\mathcal{T}$&$\mathcal{M}$&$\tilde{\mathcal{M}}$
&&\multicolumn{1}{|c|}{Dipolar-Octupolar $(n=3)$}&$\mathcal{I}$&$\mathcal{T}$&$\mathcal{M}$&$\tilde{\mathcal{M}}$\\
\hline\hline

\parbox[t]{2mm}{\multirow{5}{*}{\rotatebox[origin=c]{90}{NN}}}&$J_{zz},J_{\pm},J_{++},J_{z+}$&1&1&1&1&&$J_{zz},J_{\pm},J_{++}$&1&1&1&1&&$J_{zz},J_{\pm},J_{++},J_{z+}$&1&1&1&1\\

&$J_{--},J_{z-}$&1&1&-1&1&&$J_{z+}$&1&-1&1&1&&$J_{--},J_{z-}$&1&1&-1&1\\
&$h_z$&1&-1&-1&1&&$J_{--}$&1&1&-1&1&&$h_z,h_x$&1&-1&-1&1\\
&$\kappa_{zz},\kappa_{\pm},\kappa_{++},\kappa_{--},$&\multirow{2}{*}{-1}&\multirow{2}{*}{1}&\multirow{2}{*}{1}&\multirow{2}{*}{1}&   &$J_{z-},h_z$&1&-1&-1&1&&$h_y$&1&-1&1&1\\
&$\kappa_{z+},\kappa_{z-}$&&&&&&$\kappa_{zz},\kappa_{\pm},\kappa_{++},\kappa_{--}$&\multirow{2}{*}{-1}&\multirow{2}{*}{1}&\multirow{2}{*}{1}&\multirow{2}{*}{1}&&$\kappa_{zz},\kappa_{\pm},\kappa_{++},\kappa_{--}$&\multirow{2}{*}{-1}&\multirow{2}{*}{1}&\multirow{2}{*}{1}&\multirow{2}{*}{1}\\
&&&&&&& $\kappa_{z+},\kappa_{z-}$&&&&&&$\kappa_{z+},\kappa_{z-}$&&&&\\

\hline\hline

\parbox[t]{2mm}{\multirow{8}{*}{\rotatebox[origin=c]{90}{NNN three-spin}}}
&$t_{\pm +}$&-1&-1&1&-1&&$t_{\pm +}$&-1&1&1&-1&&$t_{\pm +}$&-1&-1&1&-1\\

&$t_{\pm z}$&1&-1&1&-1&&$t_{\pm z}$&1&-1&1&-1&&$t_{\pm z}$&1&-1&1&-1\\
&$t_{\mp -},t_{\mp z}$&1&-1&1&1&&$t_{\mp -}$&1&1&1&1&&$t_{\mp -},t_{\mp z}$&1&-1&1&1\\
&$\chi_{\pm z},\chi_{\mp z}$&1&1&-1&-1&&$t_{\mp z}$&1&-1&1&1&&$\chi_{\pm z},\chi_{\mp z}$&1&1&-1&-1\\
&$\chi_{\pm+},\chi_{\mp-}$&-1&1&-1&-1&&$\chi_{\pm z},\chi_{\mp z}$&1&1&-1&-1&&$\chi_{\pm+},\chi_{\mp-}$&-1&1&-1&-1\\
&$\kappa_{\pm z},\kappa_{\mp z}$&-1&1&1&1&&$\chi_{\pm+},\chi_{\mp-}$&-1&1&-1&-1&&$\kappa_{\pm z},\kappa_{\mp z}$&-1&1&1&1\\
&$\xi_{\pm z},\xi_{\mp z}$&-1&1&-1&-1&&$\kappa_{\pm z},\kappa_{\mp z}$&-1&1&1&1&&$\xi_{\pm z},\xi_{\mp z}$&-1&1&-1&-1\\
&&&&&&&$\xi_{\pm z},\xi_{\mp z}$&-1&1&-1&-1&&&&&&\\

\hline\hline

\parbox[t]{2mm}{\multirow{7}{*}{\rotatebox[origin=c]{90}{NNN two-spin}}}
&$J_{2++},J_{2z+}$&1&1&1&1&&$J_{2++}$&1&1&1&1&&$J_{2++},J_{2z+}$&1&1&1&1\\
&$J_{2--}$&-1&1&1&-1&&$J_{2z+}$&1&-1&1&1&&$J_{2--}$&-1&1&1&-1\\
&$J_{2z-}$&1&1&1&-1&&$J_{2--}$&-1&1&1&-1&&$J_{2z-}$&1&1&1&-1\\
&$\chi_{2z+},\chi_{2z-}$&1&1&-1&-1&&$J_{2z-}$&1&-1&1&-1&&$\chi_{2z+},\chi_{2z-}$&1&1&-1&-1\\
&$\chi_{2++},\chi_{2--}$&-1&1&-1&-1&&$\chi_{2z+},\chi_{2z-}$&1&1&-1&-1&&$\chi_{2++},\chi_{2--}$&-1&1&-1&-1\\
&$\kappa_{2z+},\kappa_{2z-}$&-1&1&1&1&&$\chi_{2++},\chi_{2--}$&-1&1&-1&-1&&$\kappa_{2z+},\kappa_{2z-}$&-1&1&1&1\\
&$\xi_{2z+},\xi_{2z-}$&-1&1&-1&-1&&$\kappa_{2z+},\kappa_{2z-}$&-1&1&1&1&&$\xi_{2z+},\xi_{2z-}$&-1&1&-1&-1\\
&&&&&&&$\xi_{2z+},\xi_{2z-}$&-1&1&-1&-1&&&&&&\\

\hline
\end{tabular}
\caption{Action of inversion $(\mathcal{I})$, time-reversal $(\mathcal{T})$, the improper symmetries of the achiral tetrahedral point group $(\mathcal{M})$ and the modified mirror symmetry ($\tilde{\mathcal{M}}$, defined in \cref{eq:define_modified_mirror_symmetry}) on the symmetry-diagonalized parameters of the Hamiltonian for the different classes of spin doublets. The rows of the table are grouped by the Hamiltonian that they appear in (see  \cref{eq:H_class3_2spin_A4,eq:H3_class3_A4,eq:H2_class3_NNN_A4,eq:H_class1_2spin_A4,eq:H3_class1_A4,eq:H2_class1_NNN_A4}  and \cref{fig:nnn_rl_parameter_description}). The table shows whether or not a sign change is required for each parameter to leave the Hamiltonian invariant when the corresponding symmetry action is applied to the spin operators in the Hamiltonian. 
Note that all the parameters have the same symmetry action under $\mathcal{I}$, $\mathcal{M}$ and $\tilde{\mathcal{M}}$ in all doublet classes. Their action under $\mathcal{T}$ is distinct for $n=2$ (see \cref{eqs:doublet_spin_symmetry_transformations}). This results in the tables for $n=1$ and $n=3$ differing only in the parameters $h_x$ and $h_y$, since these are disallowed in the class $n=1$.
} 
\label{tab:symmetry_action}

\end{table*}


In the limit where $J_{zz}$ is much larger than all the other parameters, the perturbative low-energy expansion of $H=H_0+H_3$ generates the seven-spin terms in \cref{eq:ham_yterm} with the coefficient: 
\begin{align}
&\quad Y = Y_{\pm +}+ Y_{\pm z}  
\label{eq:eb_coeff_Y_from_t}\\
&Y_{\pm +}=\frac{6\ h_z\ \text{Im} \left( \mathbf{t}_{\pm+,r} \ \mathbf{t}_{\pm+,l}^*  \right)} { J_{zz}^2} 
\label{eq:coeff_Y_pmp}\\
&Y_{\pm z} = 3\left( \frac{\left( t_{\mp z,r,B}+t_{\mp z,l,B} \right) \left(J_{\pm,B} \right)^2} {\left(J_{zz,A} \right)^2} - \frac{\left( t_{\mp z,r,A}+t_{\mp z,l,A} \right) \left(J_{\pm,A} \right)^2} {\left(J_{zz,B} \right)^2} \right)
\label{eq:coeff_Y_pmz}
\end{align} 
The full derivation of the above expressions can be found in \cref{sec:nnn_terms_and_generating_eb}. 

The expression for $Y_{\pm +}$ can be rewritten in the form:
\begin{align}
    Y_{\pm +}&=\frac{12\ h_z\   t_{\pm+} t_{\mp-} \left( 1-\chi_{\pm+} \chi_{\mp-} \right) } { J_{zz}^2}.
    \label{eq:coeff_Y_pmp_in_chi}
\end{align}
If all the breathing parameters are set to be the same (i.e. $\kappa_s=\kappa \  \forall s$), then the expression for $Y_{\pm z}$ can be simplified to
\begin{align}
Y_{\pm z}&=- 12\ \frac{t_{\mp z} J_{\pm}^2 \ }{J_{zz}^2 }  \frac{\left( \chi_{\mp z}\ \kappa\ (5+10\kappa^2+\kappa^4) +  \xi_{\mp z}\  (1+6\kappa^2+\kappa^4) \right)}{(1-\kappa^2)^2} \nonumber \\
&= - 12\ \frac{t_{\mp z} J_{\pm}^2 }{J_{zz}^2} \left(  5\,\chi_{\mp z}\ \kappa+\xi_{\mp z} \right) +O(\kappa^2).
\label{eq:coeff_Y_pmz_single_kappa}
\end{align}
Notice that the product of parameters in all terms of  \cref{eq:coeff_Y_pmz_single_kappa,eq:coeff_Y_pmp_in_chi} are odd under improper symmetries $\mathcal{I}$, $\mathcal{T}$, $\mathcal{M}$ and $\tilde{\mathcal{M}}$ (refer \cref{tab:symmetry_action}).

\subsection{Long-range two-spin terms that generate $\theta$}
\label{sec:long_range_two_spin_terms_that_generate_theta}

In this section, we show that NNN two-spin terms share symmetry properties with the three-spin terms in the last section and can generate the $Y$-term in the Schrieffer-Wolff expansion. 
Crucially, the NNN $S^z S^+$-terms and $S^+S^+$-terms upon breaking $\mathcal{I}$, $\mathcal{T}$ and $\mathcal{M}$ symmetries decompose into right and left-oriented terms similar to the three-spin terms in \cref{eq:H3_class3_A4}.

With broken $\mathcal{T}$ and $\mathcal{I}$, but retaining the full $T_d$ point group, the following NNN two-spin terms can be included:
\begin{align}
H_{2,T_d}&= \sum_{\sigma\in\{A,B\}} \sum\limits_{\braket{ijk}_{\sigma} }
\left( J_{2z+,\sigma}+i\,J_{2z-,\sigma} \right) S_i^z  S_{k}^+ 
 +\left( J_{2++}+i\,J_{2--} \right) \sum\limits_{\braket{ijk}_A}  S_i^+  S_{k}^+ 
+{\textrm{h.c.}} 
\end{align}
When the $T_d$ point group is reduced to the $T$, we obtain 12 distinct parameters:
\begin{align}
H_{2}&= \sum_{\sigma\in\{A,B\}}  \sum\limits_{\braket{ijk}_{\sigma,R} } 
\left(J_{2z+,r,\sigma}+i\,J_{2z-,r,\sigma}\right)\ S_i^z S_{k}^+   +\sum_{\sigma\in\{A,B\}} \sum\limits_{\braket{ijk'}_{\sigma,L} }\left(J_{2z+,l,\sigma} +i\,J_{2z-,l,\sigma} \right)\ S_i^z S_{k'}^+ \nonumber \\  
& + \left(J_{2++,r}+i\,J_{2--,r} \right) \sum\limits_{\braket{ijk}_{A,R}}  S_i^+ S_{k}^+  +\left(J_{2++,l} + i\,J_{2--,l}\right) \sum\limits_{\braket{ijk'}_{A,L}} S_i^+ S_{k'}^+  +  {\textrm{h.c.}} 
\label{eq:H2_class3_NNN_A4}
\end{align}
We proceed with a reparametrization akin to that applied in the previous section. 
We define $\kappa,\,\xi,\, \chi$ and bold complex parameters from the NNN parameters in the above equation similar to how they are defined in \cref{eq:define_kappa_xi,eq:define_chi,eq:define_bold_complex_parameters}. The symmetry analysis of these parameters is shown in \cref{tab:symmetry_action}.

The perturbative low energy expansion of $H=H_0 + H_2$ generates the term in \cref{eq:ham_yterm} with the coefficient
\begin{align}
    Y&=Y_{2++}+Y_{2z+}\\
    Y_{2z+} &=  C(\kappa_{\pm},\kappa_{zz})\frac{J_{\pm}^2}{4J_{zz}^3}   \left(h_y (J_{2z+,r,A} -J_{2z+,l,A}-J_{2z+,r,B} +J_{2z+,l,B})- h_x(J_{2z-,r,A} -J_{2z-,l,A}+J_{2z-,r,B} -J_{2z-,l,B})   \right)
    \label{eq:coeff_Y_2zp}\\
    Y_{2++}&= \left( 18\ \frac{(1+\kappa_{\pm}\kappa_{zz})}{1-\kappa_{zz}^2}+3 \right)\frac{h_z\ J_{\pm}\ \text{Im} \left( \mathbf{J}_{2++,r} \ \mathbf{J}_{2++,l}^*  \right)} {\left(J_{zz}\right)^3} 
    \label{eq:coeff_Y_2pp}
\end{align}
where,
\begin{align}
C(\kappa_{\pm},\kappa_{zz})&= 4\left( 2+\frac{7+6\kappa_{\pm}\kappa_{zz}-\kappa_{\pm}^2}{1-\kappa_{zz}^2}   +\frac{6\left( (1+\kappa_{zz}^2) (1+\kappa_{\pm}^2)+4\kappa_{\pm}\kappa_{zz}\right)}{(1-\kappa_{zz}^2)^2 }  \right) \label{eq:define_C_kpm_kzz}\\
&= 20(3+\kappa_{\pm}^2+5\kappa_{zz}^2+6\kappa_{\pm}\kappa_{zz})+O(\kappa^3)
\end{align}
The expression for $Y_{2z+}$ in \cref{eq:coeff_Y_2zp} assumes that $J_{\pm}>> J_{2++}, J_{2--}$. If $J_{2++}$ and $J_{2--}$ are of the same order as $J_{\pm}$, there will be additional terms in \cref{eq:coeff_Y_2zp}. The derivation of the above expressions are included in \cref{sec:nnn_terms_and_generating_eb}. 

$Y_{2z+}$ and $Y_{2++}$ can be expressed in terms of the symmetry-diagonalized parameters as follows:
\begin{align}
    Y_{2z+}&=\frac{60J_{\pm}^2  }{J_{zz}^3} \left(J_{2z+}\ h_y\left( \kappa_{2z+}\ \chi_{2z+} +\xi_{2z+} \right) - J_{2z-}\  h_x \left( \kappa_{2z-} +\xi_{2z-}\ \chi_{2z-} \right) \right) +O(\kappa^2)
    \label{eq:coeff_Y_2zp_in_chi}\\
     Y_{2++}&= \frac{42\  h_z\ J_{\pm}\ J_{2++}J_{2--} (1- \chi_{2++}\chi_{2--})} {\left(J_{zz}\right)^3} +O(\kappa^2)
     \label{eq:coeff_Y_2pp_in_chi}
\end{align}
Similar to expressions for $Y_{\pm z}$ and $Y_{\pm+}$, the product of parameters in \cref{eq:coeff_Y_2zp_in_chi,eq:coeff_Y_2pp_in_chi} break all the symmetries required to obtain a $Y$-term (see \cref{tab:symmetry_action}).

\subsection{Effective spin 1/2 and non-Kramers doublets}
\label{sec:effective_spin_half_and_non_kramers_doublets}

In a dipolar-octupolar QSI, the Hamiltonian that preserves the $T_d$ point group along with $\mathcal{I}$ and $\mathcal{T}$ is given in \cref{eq:H_class3_S4}. The corresponding Hamiltonian for QSI with effective spin 1/2 doublets ($n=1$) is
\begin{align}
H_{T_d}=\sum_{\braket{ ij}} &\left\{ J_{zz} S^z_i S^z_j - J_\pm \left(S^+_i S^-_j + S^-_i S^+_j) \right. + \left( J_{++} \  \gamma_{ij} S^+_i S^+_j + J_{z\pm}\  \zeta_{ij} (S^z_i S^+_j + S^+_i S^z_j)+ \textrm{h.c.} \right)   \right\} ,
\end{align}
where we define
\begin{align}
\gamma=\begin{bmatrix}
0&1&\omega&\omega^2\\
1&0&\omega^2 &\omega\\
\omega&\omega^2&0&1\\
\omega^2 &\omega&1&0\\
\end{bmatrix},
\label{eq:gamma_matrix}
\end{align}
with $\omega =e^{i\frac{2\pi}{3}}$ and $\zeta=-\gamma^*$.
In QSI with non-Kramers doublets ($n=2$), the Hamiltonian has the same form as above, with one distinction: $\mathcal{T}$ forces $J_{z\pm}=0$.

With $\mathcal{I}$, $\mathcal{T}$ and $\mathcal{M}$ broken and preserving just tetrahedral ($T$) point group symmetry and translational symmetry, the most general Hamiltonian for QSI with effective spin 1/2 or non-Kramers doublets ($n=1$ or $2$) is:
\begin{align}
H_0=\sum_{\sigma\in\{A,B\}}\sum_{\braket{ ij}_\sigma} \bigg\{& J_{zz,\sigma}\, S^z_i S^z_j - J_{\pm,\sigma}\,(S^+_i S^-_j + S^-_i S^+_j) + \left( \left(J_{++,\sigma} + iJ_{--,\sigma}\right)\gamma_{ij} S^+_i S^+_j + \textrm{h.c.} \right)  \nonumber \\ 
& + \left(\left(J_{z+,\sigma} + i\, J_{z-,\sigma} \right)  \zeta_{ij} (S^z_i S^+_j + S^+_i S^z_j) + \textrm{h.c.} \right)   \bigg\} +\sum_{i} h_z\ S^z_i
\label{eq:H_class1_2spin_A4}
\end{align}
This Hamiltonian has 13 real parameters.
The effective action of $\mathcal{I}$, $\mathcal{T}$, $\mathcal{M}$ and $\tilde{\mathcal{M}}$ symmetries on these parameters is summarized in \cref{tab:symmetry_action}.

The NNN three-spin terms for these doublet classes take the form:
\begin{align}
H_3= 
&\sum_{\sigma\in\{A,B\}}\sum_{\braket{ijk}_{\sigma,R} }\left(t_{\pm z,r,\sigma} + i\, t_{\mp z,r,\sigma} \right)  S_i^+ S_j^- S_{k}^z 
- \sum_{\sigma\in\{A,B\}}\sum_{\braket{ijk'}_{\sigma,L} }\left(t_{\pm z,l,\sigma} + i\, t_{\mp z,l,\sigma} \right)    S_i^+ S_j^- S_{k'}^z 
\nonumber \\
&+  \left(t_{\pm +,r}+ i\, t_{\mp -,r}\right)  \sum_{\braket{ijk}_{A,R}}  \gamma_{ij}  S_i^+ S_j^- S_{k}^+ 
- \left( t_{\pm +,l}+i\, t_{\mp -,l}\right)  \sum_{\braket{ijk'}_{A,L}} \gamma_{ij}  S_i^+ S_j^- S_{k'}^+  
+ \textrm{h.c.} 
\label{eq:H3_class1_A4}
\end{align}
Since the only difference between the above Hamiltonian and \cref{eq:H3_class3_A4} are the phases $\gamma_{ij}$ for the second set of terms, the expression for $Y_{\pm z}$ remains the same as in the case $n=3$ (\cref{eq:coeff_Y_pmp}), while the expression for $Y_{\pm +}$ gets modified to:
\begin{align}
    Y_{\pm +} &= \frac{6\ h_z\ \text{Im} \left( \mathbf{t}_{\pm+,r} \ \mathbf{t}_{\pm+,l}^* \ e^{i \frac{2\pi}{3}}  \right)} { J_{zz}^2} \nonumber\\
    &= \frac{3\ h_z\ } { J_{zz}^2} \left(2\, t_{\pm+}t_{\mp-}(1-\chi_{\pm+}\chi_{\mp-})+\sqrt{3}\, t_{\pm+}^2 (1-\chi_{\pm+}^2)-\sqrt{3}\, t_{\mp-}^2 (1-\chi_{\mp-}^2)\right)
\end{align}

The NNN two-spin terms are:
\begin{align}
H_{2}&=  \sum_{\sigma\in\{A,B\}} \sum\limits_{\braket{ijk}_{\sigma,R} } 
\left(J_{2z+,r,\sigma}+i\, J_{2z-,r,\sigma}\right)\ \gamma_{ij} S_i^z S_{k}^+  
+\sum_{\sigma\in\{A,B\}} \sum\limits_{\braket{ijk'}_{\sigma,L} }\left(J_{2z+,l,\sigma} +i\,J_{2z-,l,\sigma} \right)\ \gamma_{ij} S_i^z S_{k'}^+ \nonumber \\  
& + \left(J_{2++,r}+i\,J_{2--,r} \right) \sum\limits_{\braket{ijk}_{A,R}}  \gamma_{ij} S_i^+ S_{k}^+  +\left(J_{2++,l} + i\,J_{2--,l}\right) \sum\limits_{\braket{ijk'}_{A,L}} \gamma_{ij}  S_i^+ S_{k'}^+  +  {\textrm{h.c.}} 
\label{eq:H2_class1_NNN_A4}
\end{align}
In the case of these two classes, there is no analog of the $Y_{2z+}$ term in \cref{eq:coeff_Y_2zp}, as the tetrahedral rotational symmetries disallow the terms corresponding to $h_x$ and $h_y$. 
The analog of the $Y_{2++}$ term takes the form:
\begin{align}
    Y_{2++}&= \left( 18\ \frac{(1+\kappa_{\pm}\kappa_{zz})}{1-\kappa_{zz}^2}+3 \right)\frac{h_z\ J_{\pm}\ \text{Im} \left( \mathbf{J}_{2++,r} \ \mathbf{J}_{2++,l}^* \ e^{i \frac{2\pi}{3}}\right)} {\left(J_{zz}\right)^3}
\end{align}


\section{Contributions to $\theta'$ from gapped spinons}
\label{sec:theta_from_effective_spinon_H}

QSI in the Coulomb spin liquid phase has gapped $\mathbf{e}$-charges along with gapless photonic modes. 
In \cref{sec:lgt_and_its_continuum}, we showed that ${\theta'}$ is expected to scale as $Y/g\, f\left(Y/g \right)$ in the limit $\Gamma\to \infty$ for the simple $g$-$Y$ model (\cref{eq:H_emergent_e_and_a_ring}). 
However, the gapped $\mathbf{e}$-charges or \textit{spinons}, also contribute to the value of ${\theta'}$.
When we add an effective gapped spinon Hamiltonian coupled to the emergent electromagnetic fields, with the same symmetry and scaling arguments as before, we expect the secondary contribution due to the spinons to scale as:
\begin{align}
    {\theta'}_{\text{spinon}}= \frac{Y}{\Delta} F \left(\frac{Y}{\Delta},\frac{g}{\Delta}\right)
    \label{eq:theta_spinon_scaling}
\end{align}
where $\Delta$ is the gap of the spinons and the function, $F$, is an even function of $Y/\Delta$. 
From our numerics (see \cref{fig:theta_gmft_parametric_plots}), we find that for $Y\ll g\ll \Delta$, the function $F \left(\frac{Y}{\Delta},\frac{g}{\Delta}\right)$ is a constant.
For the pyrochlore QSI Hamiltonian we consider (where $\Delta\propto J_{zz}$), in the limit $J_{zz}\gg J_{\pm}$, we expect $\theta'_{\text{spinon}}$ to be a small correction to the value of $\theta'$ obtained from \cref{eq:theta_in_Y_Gamma}.

Obtaining an effective spinon Hamiltonian from the QSI spin Hamiltonian involves making a couple of approximations. 
We use gauge mean-field theory (gMFT) to separate the spinon rotor degrees of freedom from the gauge spin degrees of freedom and then use the rotor relaxation approximation to obtain a quadratic bosonic Hamiltonian for the spinons \cite{savaryCoulombicQuantumLiquids2012,leeGenericQuantumSpin2012,savaryQuantumSpinIce2016,desrochersSymmetryFractionalizationGauge2023, gingrasQuantumSpinIce2014}.
We explain the details of these procedures in the subsection that follows. 
While these approximations allow us to extract an effective spinon Hamiltonian, they also have drawbacks. Important among them is that not all symmetry-breaking spin interactions translate into symmetry-breaking interactions of the spinons.

After we obtain an effective quadratic spinon Hamiltonian, we compute ${\theta'}$ for the spinons by using the Brillouin-zone integrals in Ref.~\cite{naikSyntheticMagnetoelectricResponse2024}. These integrals are bosonic versions of the formula provided in Ref.~\cite{essinOrbitalMagnetoelectricCoupling2010} for computing ${\theta}$ in fermionic systems.
The effective spinon Hamiltonian is an example of a system in which we get non-zero contributions to the integral only from the `cross-gap' term arising from `non-minimally coupled' hopping. The details of this contribution are provided in \cref{sec:magentoelectric_response_of_bosonic_spinons}.

\subsection{gMFT and rotor relaxation}
\label{sec:gMFT_and_rotor_relaxation}

In pyrochlore quantum spin ice, the effective spins can be separated into rotor spinon and gauge spin degrees of freedom using a parton construction. This leads to a Hamiltonian in an extended space, with a gauge constraint that maps the space back to the original Hilbert space.
Using gMFT, the rotor spinon and the gauge spin degrees of freedom can be separated, with certain constraints relating the two separated Hamiltonians. 
The rotor spinon Hamiltonian cannot be analytically treated without additional approximations. 
The rotors can be approximated to be bosons in a few different ways \cite{savaryCoulombicQuantumLiquids2012,savaryQuantumSpinIce2016,haoBosonicManybodyTheory2014}. We choose to follow Ref.~\cite{savaryCoulombicQuantumLiquids2012} and perform a global rotor relaxation to obtain an effective quadratic bosonic Hamiltonian for the spinons.
However, note that the symmetries of the terms in the gMFT Hamiltonian are not necessarily the same as those before.

The parton construction separates the rotor ($\Phi$) and gauge spin ($s$) degrees of freedom by setting
\begin{align}
    S^z_{{\br},{\br}_\mu}= s^z_{{\br},{\br}_\mu} \quad \text{and} \quad  S^+_{{\br},{\br}_\mu}=\Phi_{\br}^\dagger\  s^+_{{\br},{\br}_\mu}\Phi_{{\br}_\mu}
    \label{eq:parton_construction}
\end{align}
where ${\br}\in A$ (the sites of the diamond lattice that are centers of upward facing tetrahedra) and ${\br}_\mu= {\br}+l \hat{u}_\mu \in B$. 
The constraint to relate the extended Hilbert space to the original one is set by enforcing the charge of the rotor, $Q$ (with $[Q_{\br},\Phi_{\br}]=\Phi_{\br}$), to be the same as the divergence of the spins:
\begin{align}
    &Q_{\br} = \sum_{\mu}\eta_{\br} \ s_{{\br},{\br}_\mu}^z 
    \label{eq:gauge_constraint}\\
    &\text{with } \eta_{\br}= \left\{\begin{aligned}
        1 &&\text{if } {\br} \in A    \nonumber\\
        -1 &&\text{if } {\br} \in B 
\end{aligned}\right. .
\end{align}

We use gMFT to obtain the effective spinon and gauge-spin Hamiltonians, $H=H_s^{\text{MF}}+H_\Phi^{\text{MF}}+\text{const}$, with the mean-field approximation done by the following decompositions:
\begin{align}
\Phi ^{\dagger}\Phi ss  \rightarrow & \Phi ^{\dagger}\Phi \braket{s}\braket{s} + \braket{\Phi ^{\dagger}\Phi} s\braket{s } +\braket{\Phi ^{\dagger}\Phi} \braket{s}s-2\braket{\Phi ^{\dagger}\Phi} \braket{s}\braket{s}\nonumber\\
\Phi ^{\dagger}\Phi sss  \rightarrow & \Phi ^{\dagger}\Phi \braket{s}\braket{s}\braket{s} + \braket{\Phi ^{\dagger}\Phi} s\braket{s} \braket{s} +\braket{\Phi ^{\dagger}\Phi} \braket{s}s\braket{s} +\braket{\Phi ^{\dagger}\Phi} \braket{s}\braket{s}s - 3\braket{\Phi ^{\dagger}\Phi} \braket{s}\braket{s}\braket{s}
\label{eq:gmft_separation}
\end{align}

In the phase space where we see $\theta$-response, we expect an All-In-All-Out(AIAO) order on top of the 2-In-2-Out classical ground space manifold. With this in mind, we use the AIAO ansatz for expectation values of the gauge spins
\begin{align}
\braket{s^z_{{\br},{\br}_\mu}}= \frac{1}{2} \sin(\vartheta) \quad\text{and}\quad \braket{s^+_{{\br},{\br}_\mu}} = \frac{1}{2} \cos(\vartheta) e^{i\varphi}.
\label{eq:aiao_ansatz}
\end{align}
Here, the angles $\varphi$ and $\vartheta$ parametrize the mean field ansatz for the spins.

In this section, we consider the dipolar-octupolar doublet QSI Hamiltonian $H=H_0+H_3$ (see \cref{eq:H_class3_2spin_A4} and \cref{eq:H3_class3_A4}), with all parameters not in \cref{eq:eb_coeff_Y_from_t} set to zero (i.e. $J_{++}=J_{--}=J_{z+}=J_{z-}=h_x=h_y =t_{\pm z}= 0$). 
Using the parton construction (in \cref{eq:parton_construction}) and introducing Lagrange parameter $\Lambda$ to enforce the gauge constraint (in \cref{eq:gauge_constraint}), we get the following Hamiltonian:
\begin{align}
H=&\sum_{{\br}\in \{A,B\}} \left\{ \left(\frac{1}{2}J_{zz,\sigma_{\br}}\, Q_{{\br}}^{2} - J_{\pm,\sigma_{\br}}\sum_{\mu,\nu\neq\mu}\Phi^{\dagger}_{{\br}_\mu}\Phi_{{\br}_\nu}s_{{\br},{\br}_\mu}^{-\eta_{{\br}} }s_{{\br},{\br}_\nu}^{+\eta_{{\br}} }\right) + \frac{h_z}{2} \eta_{\br}Q_{\br}
+ \Lambda \left( \eta_{\br}Q_{\br} - \sum_{\mu} \ s_{{\br},{\br}_\mu}^z \right) \right\} \nonumber\\
&+ \sum_{{\br}\in \{A,B\}} \sum_{\mu,\nu\neq\mu} \left\{ 
 i\, t_{\mp z,r,\sigma_{\br}}\,  s_{\br_\mu,\br}^+ s_{\br,\br_{\nu}}^- s_{\br_{\nu},\br_{\nu\chi}}^z\  \Phi_{\br_\mu}^\dagger\Phi_{{\br}_{\nu}} 
-  i\, t_{\mp z,l,\sigma_{\br}}\,   s_{\br_\mu,\br}^+ s_{\br,\br_{\nu}}^- s_{\br_{\nu},\br_{\nu\rho}}^z\  \Phi_{\br_\mu}^\dagger\Phi_{{\br}_{\nu}} + \textrm{h.c.}  \right\} \nonumber \\ 
&+ \sum_{{\br}\in B} \sum_{\mu,\nu\neq\mu} \left\{ 
\mathbf{t}_{\pm +,r}\,  s_{\br,\br_\mu}^+ s_{\br_\mu,\br_{\mu\nu}}^- s_{\br_{\mu\nu},\br_{\mu\nu\chi}}^+\  \Phi_{\br}^\dagger\Phi_{{\br}_{\mu\nu\chi}} 
- \mathbf{t}_{\pm +,l}\,    s_{\br,\br_\mu}^+ s_{\br_\mu,\br_{\mu\nu}}^- s_{\br_{\mu\nu},\br_{\mu\nu\rho}}^+\ \Phi_{\br}^\dagger\Phi_{{\br}_{\mu\nu\rho}} + \textrm{h.c.}  \right\} 
\label{eq:H_rotors_and_gauge_spins}
\end{align}    
where ${\br}_\mu = {\br}+\eta_{\br}\hat{u}_\mu$, ${\br}_{\mu\nu} = {\br}_\mu+\eta_{{\br}_\mu}\hat{u}_\nu$, ${\br}_{\mu\nu\chi} = {\br}_{\mu\nu}+\eta_{{\br}_{\mu\nu}}\hat{u}_\chi$ and $\sigma_{\br}$ is $A$ ($B$) if $\br \in A$ ($\br \in B$). Here and henceforth, the subscripts $\chi$ and $\rho$ are used to indicate right and left oriented three site hops, respectively, i.e.   $\text{sign}(\hat{u}_\mu\times\hat{u}_\nu\cdot\hat{u}_\chi)=+1$ and $\text{sign}(\hat{u}_\mu\times\hat{u}_\nu\cdot\hat{u}_{\rho})=-1$.

We separate the rotor and gauge spin Hamiltonian using the gMFT approximation (see \cref{eq:gmft_separation}) and the AIAO ansatz in \cref{eq:aiao_ansatz} to obtain the effective mean-field spinon Hamiltonian:
\begin{align}
H^{\text{MF}}_\Phi
=&\sum_{{\br}\in \{A,B\}} \left\{ \left(\frac{1}{2}J_{zz,\sigma_{\br}}\, Q_{{\br}}^{2} - \frac{1}{4}J_{\pm,\sigma_{\br}} \cos^2\vartheta\sum_{\mu,\nu\neq\mu}\Phi^{\dagger}_{{\br}_\mu}\Phi_{{\br}_\nu}\right) + \left( \frac{h_z}{2} +\Lambda \right)  \eta_{{\br}}Q_{\br}
\right\} \nonumber\\
&+ \sum_{{\br}\in B} \sum_{\mu,\nu\neq\mu} \frac{1}{8}\cos^3 \vartheta\left\{ 
\mathbf{t}_{\pm +,r} \ e^{i\varphi}\ \Phi_{\br}^\dagger\Phi_{{\br}_{\mu\nu\chi}} 
- \mathbf{t}_{\pm +,l}  \ e^{i\varphi}\ \Phi_{\br}^\dagger\Phi_{{\br}_{\mu\nu\rho}} + \textrm{h.c.}  \right\}.
\label{eq:H_MF_rotors}
\end{align}
The effective Hamiltonian for the gauge spins is
\begin{align}
H_s^{\text{MF}} &= -  \sum_{r\in A} \vec{h}_{eff}.\sum_\mu \vec{s}_{r,r_\mu} 
\label{eq:H_MF_gauge_spins}\\
h^+_{eff} &=  \frac{1}{2} \left(  J_{\pm,A}\, I_{2,B}+ J_{\pm,B}\, I_{2,A} \right)  \cos(\vartheta) e^{-i\varphi}
- \frac{3}{4}\left( \mathbf{t}_{\pm +,r} -\mathbf{t}_{\pm +,l} \right) \left( 2I_{3}+I_{3}^* \right) \cos\vartheta\sin\vartheta  \nonumber \\ 
h_{eff}^z &=  2 \Lambda \nonumber
\end{align}
In the above equations, we have used the following definitions of the correlations:
\begin{align}
I_0&=\braket{\Phi_{{\br}}^{\dagger} \Phi_{{\br}} }\nonumber \\ 
I_{2,A}&=\sum_{\nu\neq\mu}\braket{\Phi_{{\br}}^{\dagger} \Phi_{{\br}_{\mu\nu}} }\nonumber \quad \text{with } \br\in A \nonumber \\ 
I_{2,B}&=\sum_{\nu\neq\mu}\braket{\Phi_{{\br}}^{\dagger} \Phi_{{\br}_{\mu\nu}} }\nonumber \quad \text{with } \br\in B \\ 
I_{3}&=\braket{\Phi_{r}^{\dagger} \Phi_{r_{\mu\nu\chi}} } 
\end{align}

Note that the terms with $t_{\mp z}$ coefficients have dropped out in \cref{eq:H_MF_rotors}. These terms attempt to give a pure imaginary component to the two-site hopping of the spinons. 
The tetrahedral ($T$) symmetry in the diamond lattice prevents the two site hoppings of the spinons from gaining an imaginary component. Hence, the $\theta$-response from the $t_{\mp z}$ term we saw from the perturbative analysis in \cref{eq:coeff_Y_pmz} is not reflected in the gMFT Hamiltonian for the spinons. At this point, we simplify the Hamiltonian further by setting $\kappa_{zz}=\kappa_{\pm}=0$  (which sets $J_{zz,A}=J_{zz,B}=J_{zz}$ and $J_{\pm,A}=J_{\pm,B}=J_{\pm}$), since it does not appear in \cref{eq:coeff_Y_pmp}.

We now convert the rotor Hamiltonian in \cref{eq:H_MF_rotors} to its corresponding Lagrangian, relax the rotors to complex scalars by adding a global Lagrange parameter $\lambda$, and convert the Lagrangian of the complex scalars back to its corresponding Hamiltonian to obtain the following  effective quadratic bosonic Hamiltonian for the spinons:
\begin{align}
H^{\text{MF}}_\psi=&\sum_{{\br}\in \{A,B\}} \Bigg\{ \Bigg( 
2J_{zz}\ \Pi_{{\br}}^{2} 
- \frac{1}{4}J_{\pm} \cos^2\vartheta\sum_{\mu,\nu\neq\mu}\Phi^{\dagger}_{{\br}_\mu}\Phi_{{\br}_\nu}  + \lambda \Phi_{\br}^\dagger\Phi_{\br}   \Bigg)
+i \left( \frac{h}{2} +\Lambda \right)  \eta_{{\br}} ( \Pi_{\br}^\dagger\Phi_{\br}- \Phi_{\br}^\dagger\Pi_{\br})\Bigg\} \nonumber\\
&+ \sum_{{\br}\in B} \sum_{\mu,\nu\neq\mu} \frac{1}{8} \cos^3 \vartheta\left\{ 
\mathbf{t}_{\pm +,r} \  \Phi_{\br}^\dagger\Phi_{{\br}_{\mu\nu\chi}} 
- \mathbf{t}_{\pm +,l}  \ \Phi_{\br}^\dagger\Phi_{{\br}_{\mu\nu\rho}} + \textrm{h.c.}  \right\}
\label{eq:H_MF_bosonic_spinons}
\end{align}
The complex scalar have commutation relations $[\Phi_{\br},\Pi_{\br'} ^{\dagger}]=i \delta_{\br,\br'}$ while maintaining $[Q_{\br},\Phi_{\br'}]=\Phi_{\br}\,\delta_{\br,\br'}$, with $Q_{\br}=i(\Pi_{\br} ^{\dagger}\Phi_{\br}-\Pi_{\br}\Phi_{\br} ^{\dagger})$. The parameters in the above Hamiltonian have to satisfy the following consistency conditions:
\begin{align}
Q&=2\sin\vartheta \nonumber\\
I_0&=1 \nonumber\\
\tan \vartheta&= \frac{4\Lambda}{
    2 \left(  J_{\pm,A}\, I_{2,B}+ J_{\pm,B}\, I_{2,A} \right)  \cos(\vartheta)
    - 3\left( \mathbf{t}_{\pm +,r} -\mathbf{t}_{\pm +,l} \right) \left( 2I_{3}+I_{3}^* \right)e^{i\varphi} \cos\vartheta\sin\vartheta 
}
\label{eq:MF_consistency_conditions}
\end{align}
These conditions enforce the gauge constraint, the rotor constraint, and consistency with the effective gauge spin Hamiltonian, respectively.

\subsection{Magnetoelectric response of the bosonic spinons}
\label{sec:magentoelectric_response_of_bosonic_spinons}

In this subsection, we first introduce the magnetoelectric tensor of insulating systems and describe how they are computed in bosonic lattice insulators. 
We then compute the $\theta'_{\text{spinon}}$, the isotropic component of the magnetoelectric tensor of the spinon Hamiltonian, and verify its scaling in \cref{eq:theta_spinon_scaling}.

\begin{figure}[]
    \centering
    \includegraphics[width=0.35\linewidth]{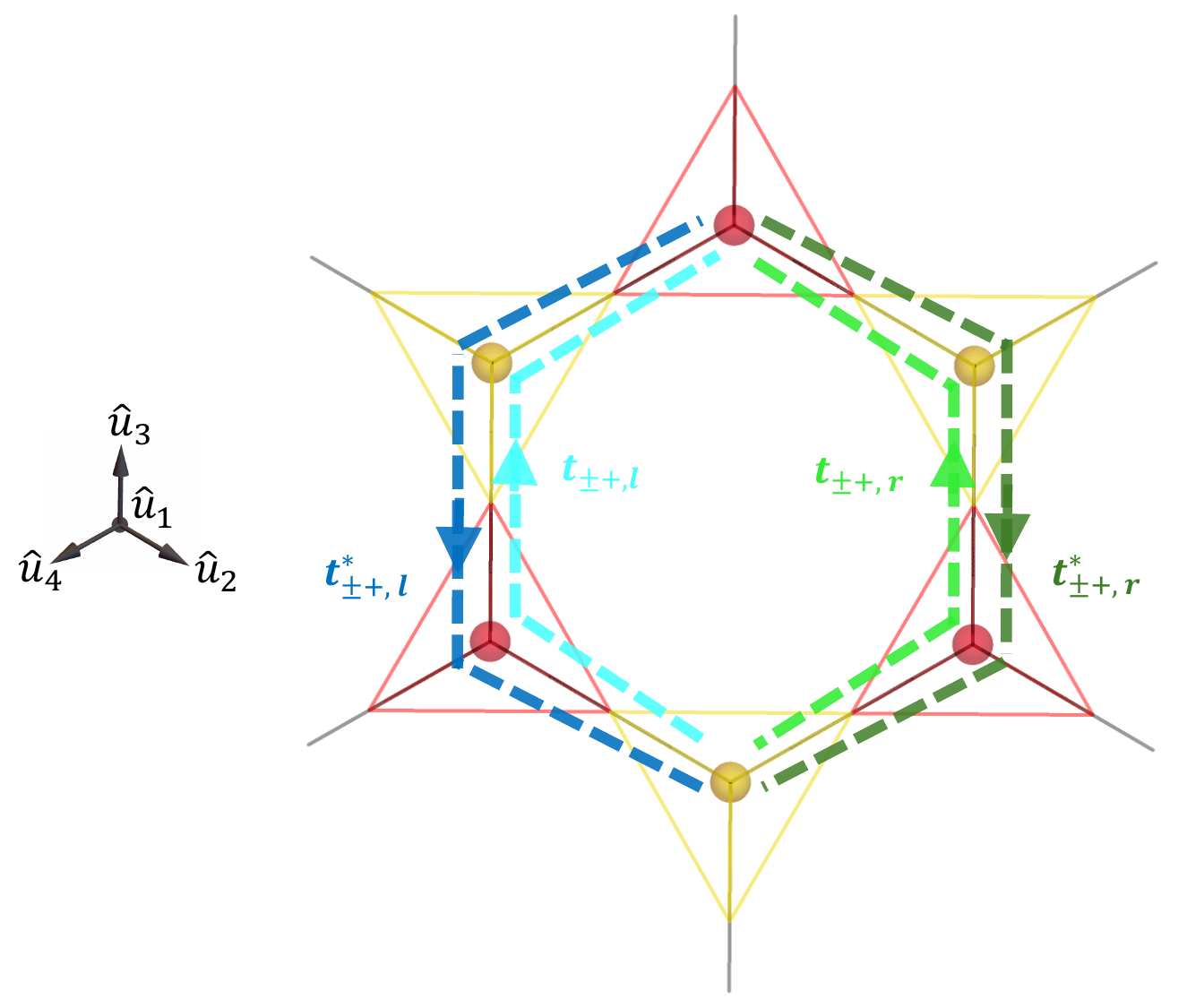}
    \caption{The spinon three-site hops on the diamond lattice. The zero-field tight-binding bosonic Hamiltonian in \cref{eq:H_MF_bosonic_spinons} connects all diamond lattice hops on the opposite ends of the hexagonal plaquettes through left and right oriented three-site hops. At zero magnetic field, a hop to and fro between such sites does not distinguish between the two orientations. However, when an external magnetic field is applied, the hops accumulate path-dependent phases, and a hop around the hexagon can accumulate an Aharonov-Bohm phase. }
    \label{fig:3_site_hops_diamond}
\end{figure}

\begin{figure*}
    \centering
    \includegraphics[width=
    \linewidth]{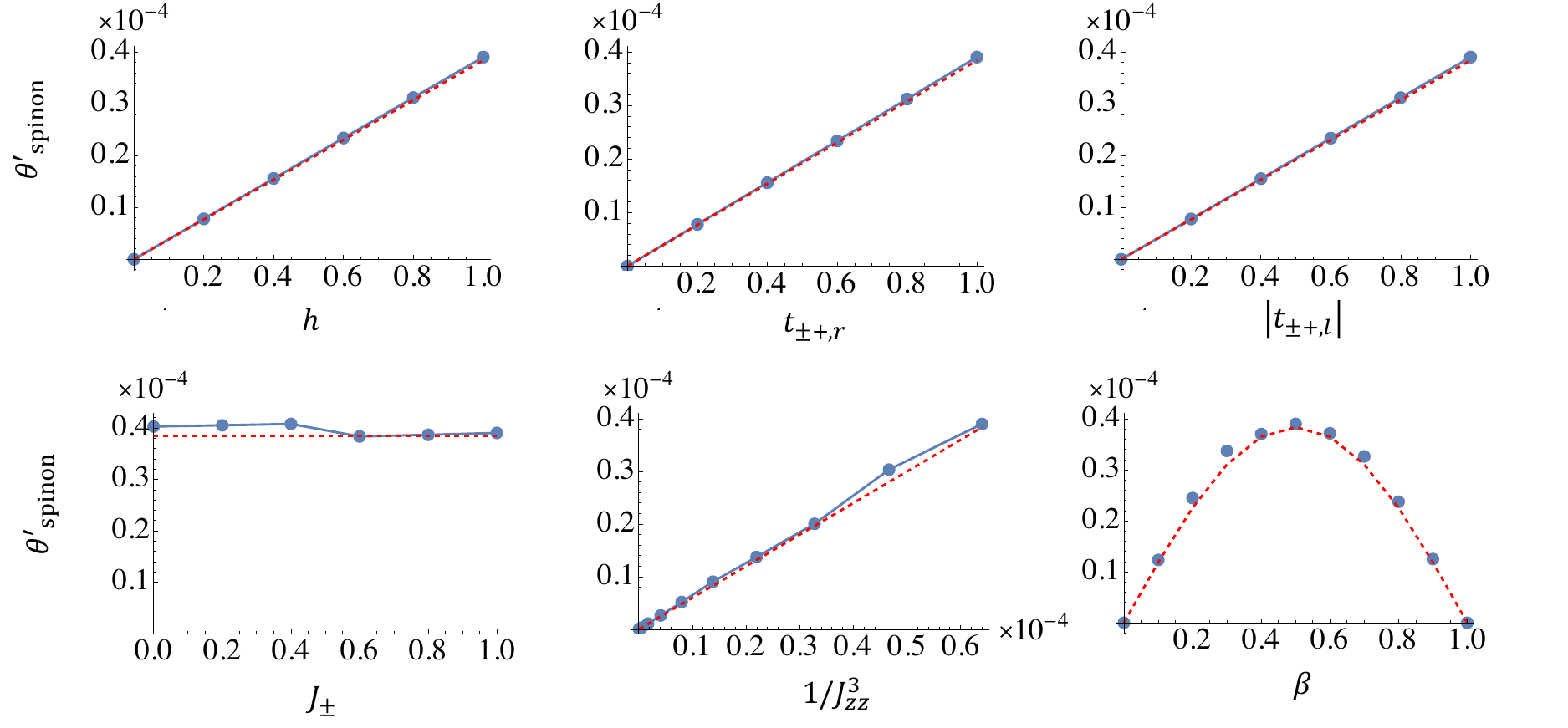}
    \caption{The parametric dependence of ${\theta'}_{\text{spinon}}$ computed by numerically performing the associated $k$-space integral in \cref{eq:theta_bz_int} using the effective spinon Hamiltonian in \cref{eq:H_MF_bosonic_spinons} is shown in blue. 
    The red dotted line shows the value of ${\theta'}$ we expect from \cref{eq:theta_spinon_scaling_in_H_parameters} with the scaling function fit to a constant, $\tilde{F} \left(\frac{Y_{\pm+}}{J_{zz}},\frac{g}{J_{zz}}\right)=0.05$.
    We choose parameters $J_{zz}=25,\, J_{\pm}=1,\, h_z=1, t_{\pm +,r}=1$ and $t_{\pm +,l}=-i$ and plot ${\theta'}_{\text{spinon}}$ while varying one of the parameters in each plot. For the last plot, we set $t_{\pm +,l}=e^{-i \beta \pi}$ and vary $\beta$.  
    The values of $\vartheta$, $\varphi$, $\lambda$ and $\Lambda$ are set by the consistency conditions in \cref{eq:MF_consistency_conditions} by computing correlations for each set of parameter values. 
    The Brillouin zone integrals are numerically evaluated using a uniform grid of size $30^3$ in the reciprocal space. (Color online.)
    }
    \label{fig:theta_gmft_parametric_plots}
\end{figure*}

Magnetoelectric response is the polarization ($\mathbf{P}$) response to external magnetic fields ($\mathbf{B}$) and the magnetization ($\mathbf{M}$) response to external electric fields ($\mathbf{E}$). The magnetoelectric tensor of a system is defined by 
\begin{align}
    \upalpha^i_{j}=\frac{\partial P^i}{\partial B^j}= \frac{\partial M_j}{\partial E_i}.
    \label{eq:alpha_defn}
\end{align}
The $\theta$-coefficient of such a system can be obtained by computing the isotropic contribution to the magnetoelectric tensor:
\begin{align}
    \upalpha^i_{\ j}= 4 \pi^2\ {\theta}\ \delta^i_j + \tilde{\upalpha}^i_j
\end{align}
where, $\tilde{\upalpha}^i_j$ is a traceless tensor giving the anisotropic contribution to the magnetoelectric tensor.
The magnetoelectric tensor of a bosonic lattice insulator can be computed by the Brillouin-zone integrals in Ref.~\cite{naikSyntheticMagnetoelectricResponse2024}. 
These integrals are similar to their fermionic counterparts \cite{essinOrbitalMagnetoelectricCoupling2010} and are derived by computing the polarization response of the bosonic system when an external magnetic field is applied. In general, the magnetoelectric tensor has a Chern-Simons contribution and cross-gap contributions,
\begin{align}
    \upalpha^i_{\ j}= \upalpha_{\text{CS}}\, \delta^i_j+ {(\upalpha_{\text{G}})}^i_j.
\end{align}
The Brillouin-zone integral formulae to compute these contributions in bosonic systems are in terms of the dynamical matrix, $D$, which is the matrix that sets the equations of motion,
\begin{align}
    i\partial_t\begin{pmatrix} \Pi_{\bm{r}} \\
    \Phi_{\bm{r}} \end{pmatrix} 
    = \sum_{\br,\br'} D_{\br,\br'} \begin{pmatrix} \Pi_{\bm{r'}} \\
    \Phi_{\bm{r'}} \end{pmatrix}.
\end{align}
The dynamical matrix can be obtained from the Hamiltonian of the bosonic system and can be broken into two parts
\begin{align}
    D=D_0+D'
\end{align}
where, in $D_0$, all the hopping terms are minimally coupled to the magnetic field via the Peierl's phase along the shortest path between the corresponding lattice points, and $D'$ contains information about all the other ways the magnetic field couples to the system. Although $D'=0$ at zero magnetic field, $\mathbf{B}=0$, its derivative $\partial D' / \partial B^i$ at zero magnetic field, which is part of the expression for $\upalpha_{\text{G}}$, can be non-zero.

To compute the contribution to $\theta'$ from the spinons, we need to compute the magnetoelectric response of the spinons of the QSI system to the emergent electromagnetic fields of the system. 
For the spinon-hopping Hamiltonian in \cref{eq:H_MF_bosonic_spinons}, the coupling to the emergent magnetic field is not through just a Peierl's phase minimal coupling (i.e. through the shortest Euclidean path between the two sites), but rather a Peierl's phase coupling through the shortest path along the pyrochlore lattice:
\begin{align}
    H_{\psi,\text{hop}}^{\text{MF}}= -&\sum_{{\br}\in \{A,B\}}  
 \frac{1}{4}J_{\pm} \cos^2\vartheta\sum_{\mu,\nu\neq\mu} e^{i\left(\int_{\br_\nu}^\br \mathbf{a}.d\mathbf{x}+\int_{\br}^{\br_\mu} \mathbf{a}.d\mathbf{x}\right)}\Phi^{\dagger}_{{\br}_\mu}\Phi_{{\br}_\nu} 
 \nonumber\\
&+ \sum_{{\br}\in B} \sum_{\mu,\nu\neq\mu} \frac{1}{8} \cos^3 \vartheta  \left\{  e^{i\left(\int_{\br_{\mu\nu}}^{\br_\mu} \mathbf{a}.d\mathbf{x}+\int_{\br_\mu}^{\br} \mathbf{a}.d\mathbf{x}\right)} \left(  
e^{i\int_{\br_{\mu\nu\chi}}^{\br_{\mu\nu}} \mathbf{a}.d\mathbf{x}} \mathbf{t}_{\pm +,r} \  \Phi_{\br}^\dagger\Phi_{{\br}_{\mu\nu\chi}} 
- e^{i\int_{\br_{\mu\nu\rho}}^{\br_{\mu\nu}} \mathbf{a}.d\mathbf{x}} \mathbf{t}_{\pm +,l}  \ \Phi_{\br}^\dagger\Phi_{{\br}_{\mu\nu\rho}}  \right) + \textrm{h.c.}  \right\}
\end{align}
where $\mathbf{a}(\mathbf{x})$ is the vector potential and its curl, $\nabla\times\mathbf{a}=\mathbf{b}$, is the emergent magnetic field.
Our spinon system (with the Hamiltonian in \cref{eq:H_MF_bosonic_spinons}) is isotropic (since it has a tetrahedral point group symmetry) and gets a non-zero contribution only from one of the two cross-gap terms:
\begin{align}
{\theta'}_{\text{spinon}} &= \frac{8\pi^2}{3} \int_{\text{BZ}} \frac{d^3{k}}{(2\pi)^3} \sum_{n, m} \frac{ \sum_{i=1}^3 \text{Im} \ \text{Tr} \ P_{-n} \  \partial^i P P_{+m} (\partial D'/\partial b^i) }{E_{-n} + E_{+m}}
\label{eq:theta_bz_int}
\end{align}
Here, $D'=D-D_0$ is the `non minimally-coupled' part of the dynamical matrix obtained from the Hamiltonian in \cref{eq:H_MF_bosonic_spinons}, $\mathbf{b}$ is the emergent magnetic field that the spinons couple to, $P$ is the projector onto the space of all negatively charged annihilation modes of the bosonic system, $P_m$ is the projector onto a specific mode $m$, $E_m$ is the energy of mode $m$, and sum over index $n$ ($m$) runs over all negatively (positively) charged annihilation modes. See Ref.~\cite{naikSyntheticMagnetoelectricResponse2024} for a complete description of the projectors.

To understand why this is the only non-zero contribution to the magnetoelectric tensor, we need to turn to the subtle way in which  $\mathcal{T}$-symmetry is broken in the system. In the bosonic system, time reversal is defined by the transformations: $\Phi\to \Phi, \Pi \to -\Pi, i\to -i$. If we look at the bosonic Hamiltonian in \cref{eq:H_MF_bosonic_spinons} without considering where it came from, we would conclude it does not break $\mathcal{T}$-symmetry. 
To notice that it does, we need to recognize that this model does indeed have non-trivial Aharanov-Bohm fluxes. In \cref{fig:3_site_hops_diamond}, note that the spinon does not retrace its path during the hop with strength $\mathbf{t}_{\pm+,r}$ and that with strength $\mathbf{t}_{\pm+,l}^*$. 
This is reflected in the Hamiltonian only when a non-zero magnetic field, $\mathbf{b}\neq0$, is applied: the phase $\int_P \mathbf{a}(\mathbf{x})\cdot d\mathbf{x}$ that couples to the hopping terms is path ($P$) dependent.
In the $k$-space integrals, $D_0$ of the system is oblivious to this path dependence and the breaking of $\mathcal{T}$-symmetry. Hence, the Chern-Simons and the cross-gap contribution dependent on $D_0$ are zero. 
The path dependence is encoded in $D'$, and the contribution to the magnetoelectric tensor shown in \cref{eq:theta_bz_int} is the only contribution sensitive to this. 

The results from numerically calculating the Brilluoin-zone integrals in \cref{eq:theta_bz_int} are shown in \cref{fig:theta_gmft_parametric_plots}. 
With the spinon gap $\Delta=J_{zz}/2$ and  $Y=Y_{\pm +}/64$ (see \cref{eq:H_Y_emergent_e_and_a,eq:coeff_Y_pmp}) obtained from our perturbative analysis, we expect ${\theta'}_{\text{spinon}}$ to scale as (from \cref{eq:theta_spinon_scaling}): 
\begin{align}
    {\theta'}_{\text{spinon}}= - \tilde{F} \left(\frac{Y_{\pm+}}{J_{zz}},\frac{g}{J_{zz}}\right) \frac{3\ h_z\ \text{Im} \left( \mathbf{t}_{\pm+,l} \ \mathbf{t}_{\pm+,r}^*  \right)} { 16\, J_{zz}^3} 
\label{eq:theta_spinon_scaling_in_H_parameters}
\end{align}
When $J_{zz}$ is much greater than all the other parameters of the Hamiltonian, $\tilde{F} \left(\frac{Y_{\pm+}}{J_{zz}},\frac{g}{J_{zz}}\right)$ is expected to be a constant. For the parameters chosen in \cref{fig:theta_gmft_parametric_plots}, we find the constant to be $0.05$. 
To construct this plot, for each set of Hamiltonian parameters, we self consistently choose a value of $\lambda$ which sets $I_0=1\pm0.01$. 
For these values, we find that $|Q_{\br}|<10^{-4}$, which sets the magnitude of $\vartheta,\Lambda$ and $\varphi$ to be of the order $10^{-5}$ (see consistency conditions in \cref{eq:MF_consistency_conditions}). This allows us to set these three parameters to zero without introducing significant errors in computing the integrals in \cref{eq:theta_bz_int}. The plots in \cref{fig:theta_gmft_parametric_plots} show that there is excellent agreement between the numerics with the Hamiltonian in \cref{eq:H_MF_bosonic_spinons} and the expected scaling of ${\theta'}_{\text{spinon}}$ from \cref{eq:theta_spinon_scaling_in_H_parameters}. 

While these numerics can only capture $\theta$-response arising from one of the four different contributions to $Y$, it is remarkable that numerics with the Bloch ground state correlators of an effective bosonic Hamiltonian obtained through the mean-field and rotor relaxation approximations agree so well with scaling expected through our perturbative analysis of the original Hamiltonian.


\end{widetext}

\section{Discussion}
\label{sec:discussion}

In this paper, we have determined the microscopic ingredients required for a $\theta$-term to appear in the long-wavelength emergent quantum elecrodynamics governing the deconfined phase of pyrochlore quantum spin ice. 
On a fundamental level, the local microscopic Hamiltonian must break $\mathcal{P}$, $\mathcal{T}$ and $\mathcal{M}$ symmetry to support a $\theta$-term. 
We analyze the plethora of local terms permitted by this symmetry reduction (from $F\bar{d}3m$ to $F23$) within the canonical lattice-scale spin Hamiltonians for pyrochlore spin ice.
In the regime where the ice interaction remains dominant, we show how these two- and three-body spin interactions can be Schrieffer-Wolff projected into the ice manifold to generate a minimal two-parameter ``$g-Y$'' Hamiltonian at the ring exchange scale. 
From the $g-Y$ model, it is straightforward to understand the emergence of the long-wavelength $\theta$-electromagnetism, as the $Y$-term is a natural lattice regularization of the $\theta$-term itself.
At $Y=0$, the $g-Y$ model reduces to the celebrated pyrochlore ring-exchange model, which is known to lie in the deconfined phase \cite{hermelePyrochlorePhotons$U1$2004} with an effective Maxwell action.
For small $Y/g$, this action is supplemented by a $\theta$-term (\cref{eq:action_theta}), with a coefficient that scales as $\theta'=\frac{Y}{g}\,f\left(\frac{Y}{g}\right)$. 
Our derivations allow parametric estimates of $Y$ from the microscopic spin interactions and, in turn, estimates of $\theta'$ from the $g-Y$ theory analyzed in certain limits. We also estimate contributions to $\theta'$ from the gapped spinons in the system (\cref{eq:theta_spinon_scaling}) within certain gauge mean-field approximations.

We leave to future research the precise determination of the phase diagram and long-wavelength couplings of the $g-Y$ theory, which should be amenable to numerical investigations beyond the scope of this work.

Material realizations of a Coulomb quantum spin liquid with non-zero ${\theta'}$ appear most plausible in a breathing pyrochlore compound with local AIAO fields induced by an underlying antiferromagnetic order in the material. While breathing pyrochlores \cite{kimuraExperimentalRealizationQuantum2014,rauBehaviorBreathingPyrochlore2018,bagSingleTetrahedronPhysics2023,genBreathingPyrochloreMagnet2023,okamotoBreathingPyrochloreLattice2013,sharmaSynthesisPhysicalMagnetic2022,zhangMossbauerSpectroscopyStudy2023} and pyrochlores with staggered AIAO fields \cite{jaubertMonopoleHolesPartially2015,chenMagneticMonopoleCondensation2016,petitObservationMagneticFragmentation2016,lefrancoisFragmentationSpinIce2017,cathelinFragmentedMonopoleCrystal2020,pearceMagneticMonopoleDensity2022} have been separately observed, we are not aware of any material that exhibits both features simultaneously.

If $\theta$ is a dynamical variable, rather than a static parameter, it is known as an axion field. 
Axions are studied within high energy physics, where they are commonly proposed as dark matter candidates \cite{duffyAxionsDarkMatter2009}, and are believed to explain the strong CP problem \cite{peccei$mathrmCP$ConservationPresence1977,kimAxionsStrong$CP$2010}. 
Although our discussion has been phrased solely in terms of a static $\theta'$, the symmetry requirements on an emergent axion field in pyrochlore QSI are identical \cite{paceDynamicalAxions$U1$2023} and our analysis naturally carries over. 
In order for the static $\theta'$-term we have described here to become a dynamical axion field, $Y$ in \cref{eq:ham_yterm} must become a dynamical variable. 
One natural way for this to occur is through the local fields generated by AIAO order.
\cref{eq:coeff_Y_pmp,eq:coeff_Y_2pp} shows how the staggered AIAO field, $h_z$, couples to the emergent ${\theta'}$. In iridate pyrochlore materials, these staggered fields arise from the interpenetrating lattice of Ir$^{4+}$ ions that order antiferromagnetically at comparatively high temperatures of the order of 100K \cite{lhotelFragmentationFrustratedMagnets2020} (spin ice is expected at temperatures of order 1K). Because of the large separation of energy scales, the fields $h_z$ are usually treated as spatially uniform and static in studies of spin ice. However, if one goes beyond this mean-field analysis and incorporates dynamical fields from the Ir$^{4+}$ ions, our emergent ${\theta'}$ parameter becomes dynamical, and we thus have a theory with emergent axions.

We shall end with some comments on how the emergent $\theta'$ in QSI relates to experimental observables and material response properties. 
In the dynamical (axion) case, $\theta'$ is predicted to manifest in the dynamical structure factor \cite{paceDynamicalAxions$U1$2023} and could thus be diagnosed in a neutron scattering experiment. If $\theta'$ is static it does not enter the bulk equations of the emergent Maxwell theory. One could imagine that it would impact the way QSI responds to external electric and magnetic fields. On general grounds, the joint action of the internal and external Maxwell theory is given by 
\begin{align}
\nonumber 
 S  = \int dt\, d^3 x &\left( \frac{1}{8\pi \alpha c} \left( \mathbf{E}^2-c^2 \mathbf{B}^2 \right)
+\frac{1}{8\pi \alpha' {c'}} \left(\mathbf{e}^2 -{c'}^2 \mathbf{b}^2 \right) \right. \\
&- \left. \frac{\theta'}{4\pi^2} \mathbf{e}\cdot \mathbf{b}  
- g_{1} \ \mathbf{e}\cdot \mathbf{B}- g_{2} \ \mathbf{b}\cdot \mathbf{E}\right)
\end{align}
where $g_1$ and $g_2$ are dimensionless parameters \cite{laumannHybridDyonsInverted2023}. 
For static external fields, a finite sample of QSI displays a magnetoelectric response quantified by (see \cref{app:magentoelectric_response_of_QSI})
\begin{equation}
    \theta \propto \alpha'^2 \,g_{1}g_{2}\,\theta'
    \, .
\end{equation}
This bulk response does, however, rely on all three of $g_1$, $g_2$, and $\theta'$ being non-zero. 
In the case of a simple magnetic insulator, one expects no or, at most, a very small coupling to the external electric field -- i.e., for realistic QSI candidates we can assume $g_2\approx 0$.
A significant internal $\theta'$ could thus exist for QSI in the absence of any conventional magnetoelectric response to external fields. 
This is a symptom of the distinction between $\theta$ as a response action, as found in other condensed matter settings, and the true emergent $\theta'$ we find in QSI.

\section*{Acknowledgments}
We would like to thank M. O. Flynn, S. Pace, C. Castelnovo, R. Moessner, R. Sch{\"a}fer, Y. Iqbal and B. Placke for stimulating discussions and collaborations on related work. 
CRL thanks the Max Planck Institute for the Physics of Complex Systems for its hospitality. 
This work was in part supported by the National Science Foundation through Grant No. PHY-1752727.

\appendix

\section{Lagrangian with $\theta$-term}
\label{sec:Lagrangian_with_theta_term}

We work in units where the electric charge $e=1$ and $\hbar =1$.
The free electromagnetic Lagrangian density with the $\theta$-term in these units is \cite{tongLecturesQuantumField2007,tongLecturesGaugeTheory2018}:
\begin{align}
\mathcal{L}=\frac{1}{8\pi {\alpha} c} \left( \mathbf{E}^2-c^2 \mathbf{B}^2 \right) - \frac{\theta}{4\pi^2} \mathbf{E}\cdot\mathbf{B} 
\label{eq:L_free_EM}
\end{align}
where, $\mathbf{E}=-\boldsymbol{\nabla}\phi-\dot{\mathbf{A}}$, $\mathbf{B}=\boldsymbol{\nabla}\times\mathbf{A}$, $\alpha$ is the fine-structure constant and $c$ is the speed of light.
In the Coulomb gauge ($\phi=0$ and $\boldsymbol{\nabla}\cdot\mathbf{A}=0$) the conjugate momentum is
\begin{align}
\mathcal{E}^i = \frac{\partial \mathcal{L}}{\partial \dot{A}_i}=- \frac{1}{4\pi\alpha c} E^i+ \frac{\theta}{4\pi^2}B^i
\label{eq:calE_combination_of_E_B}
\end{align}
This satisfies the commutation relation
\begin{align}
[A_i(\mathbf{x}),\mathcal{E}_j(\mathbf{x'})]=i\left( \delta_{ij}-\frac{\partial_i \partial_j}{\nabla^2} \right) \delta^3(\mathbf{x-x'}).
\end{align}
The commutation structure is altered to account for the fact that only two of the three components of $\mathbf{A}$ are independent. The constraints $\boldsymbol{\nabla}\cdot\mathbf{A}=0$ and $\boldsymbol{\nabla}\cdot\boldsymbol{\mathcal{E}}=0$ can be satisfied with the modified commutation relation above.

With the conjugate momentum, we can compute the Hamiltonian density of the system to be:
\begin{align}
\mathcal{H}&= \boldsymbol{\mathcal{E}}\cdot\dot{\mathbf{A}} - \mathcal{L}\nonumber \\ 
&=\frac{1}{8\pi\alpha c}\left( \mathbf{E}^2+c^2 \mathbf{B}^2 \right) \\ 
&=2\pi\alpha c \ \boldsymbol{\mathcal{E}}^2+\frac{1}{8\pi\alpha c}\left( 1+4\alpha^2\theta \right) c^2 \mathbf{B}^2-\frac{\alpha c}{\pi}\theta\ \boldsymbol{\mathcal{E}}.\mathbf{B}
\label{eq:H_free_field}
\end{align}

In the main text, we work with emergent fields $\varepsilon$ and $a$ living on the pyrochlore lattice sites. The free continuum analogs of these fields are $\boldsymbol{\mathcal{E}}$ and $\mathbf{A}$ respectively.


\section{Discrete sums to continuum integrals}
\label{sec:discrete_sums_to_continuum_integrals}

In this section, we calculate the continuum limit of the terms in \cref{eq:L_continuum}. The first term can be approximated as:
\begin{align}
\sum_i \dot{a}_i^2= l^2 \sum_{\mathbf{r}} \sum_{k=1}^4 \hat{u}_k^i \ \dot{\alpha}(\bar{{\mathbf{r}}}_k)^i  \ \hat{u}_k^j \ \dot{\alpha}(\bar{{\mathbf{r}}}_k)^j
\end{align}
Here, we have used ${\mathbf{r}}$ to denote the positions of all the $A$ sublattice sites. These uniquely identify the positions of the unit cells. The subscripts $k$ enumerate the four edges that originate at ${\mathbf{r}}$.
Note that
\begin{align}
\dot{\alpha}(\bar{{\mathbf{r}}}_k)^i = \dot{\alpha}({\mathbf{r}})^i + \frac{\partial \dot{\alpha}(\bar{{\mathbf{r}}})^i}{\partial r^j} (\bar{\mathbf{r}}_k - {\mathbf{r}})^j+ \dots
\end{align}
Using the relation
\begin{align}
\sum_{k=1}^4 \hat{u}_k^i \ \hat{u}_k^j =\frac{4}{3 } \delta^{ij},
\end{align}
we end up with:
\begin{align}
\sum_i \dot{a}_i^2 &= \frac{4}{3}l^2 \sum_{\mathbf{r}} \dot{\boldsymbol{\alpha}}({\mathbf{r}})^2 +O(\dot{\alpha}_i \partial_j \dot{\alpha}^i\delta r^j) \nonumber \\ 
&\approx \frac{4}{3}l^2 \int \frac{\mathrm{d}^3x}{V_{\text{uc}}} \  \dot{{\boldsymbol{\alpha}}}({\mathbf{x}})^2 
\end{align}
where $V_{\text{uc}} =4\sqrt{ 2 }\ l^3$ is the volume of the unit cell in the pyrochlore lattice.
Similarly, we can get:
\begin{align}
\sum_p (da)_p^2 \approx \frac{4}{3} \mathcal{A}_h^2\int \frac{\mathrm{d}^3x}{V_{\text{uc}}} (\boldsymbol{\nabla}\times {\boldsymbol{\alpha}}({\mathbf{x}}))^2
\end{align}
where $\mathcal{A}_h=\frac{3\sqrt{3}}{2}\ l^2$ is the area of the hexagon projected onto the Kagome place. 

The third term in the Lagrangian requires a couple of extra steps:
\begin{align}
&\sum_p \sum_{i\in D(p)} \dot{a}_i\cdot(da)_p =\sum_i \sum_{p\in D(i)} \dot{a}_i\cdot(da)_p \nonumber \\ 
&= l \mathcal{A}_h \sum_{\mathbf{r}} \sum_{k=1}^4 \hat{u}_k^i \ \dot{\alpha}(\bar{{\mathbf{r}}}_k)^i \ t_k^j \sum_{p\in D(\bar{\mathbf{r}}_k)} (\boldsymbol{\nabla}\times {\boldsymbol{\alpha}} (\bar{\mathbf{r}}_k +\delta {\mathbf{r}}_p))^j
\end{align}
We expand the term
\begin{align}
\sum_{p\in D(\bar{\mathbf{r}}_k)} (\boldsymbol{\nabla}\times {\boldsymbol{\alpha}} (\bar{\mathbf{r}}_k +\delta {\mathbf{r}}_p))^j= 6 (\boldsymbol{\nabla}\times {\boldsymbol{\alpha}} (\bar{\mathbf{r}}_k ))^j +O(\delta r)
\end{align}
and then again expand the term around the point ${\mathbf{r}}$ to get:
\begin{align}
\sum_p \sum_{i\in D(p)} \dot{a}_i\cdot (da)_p &= \frac{4}{3}l \mathcal{A}_h \sum_{\mathbf{r}} \dot{{\boldsymbol{\alpha}}}({\mathbf{r}})\cdot(6 \boldsymbol{\nabla}\times{\boldsymbol{\alpha}}({\mathbf{r}})) + \cdots \nonumber \\ 
&\approx 8 l \mathcal{A}_h \int \frac{\mathrm{d}^3x}{V_{\text{uc}}}  \dot{{\boldsymbol{\alpha}}}({\mathbf{x}})\cdot \boldsymbol{\nabla}\times{\boldsymbol{\alpha}}({\mathbf{x}}) 
\end{align}


\section{Symmetry action on the doublet classes}
\label{sec:symmetry_action_on_effective_spins}

The doublets of the pyrochlore ions can be classified into one of three symmetry classes \cite{rauFrustratedQuantumRareearth2019,laumannHybridDyonsInverted2023}: effective spin 1/2 ($n=1$), non-Kramers ($n=2$) and dipolar-octupolar ($n=3$).

The doublets of the class $n$ can be symmetry analyzed by considering the effective spin operators to be:
\begin{align}
S^z=\frac{1}{n}\sum_{i=1}^n \sigma^z_i &&S^\pm = \prod_{i=1}^n \sigma^\pm_i
\end{align}
These spins have the following Rotation($R$), time-reversal($\mathcal{T}$) and inversion ($\mathcal{I}$) symmetry transformations:
 \begin{align}
R: & S^z_{\mathbf{r}} \rightarrow |R| S^z_{R^{-1}\mathbf{r}} \nonumber\\
& S^\pm_{\mathbf{r}} \rightarrow |R|e^{\mp i n \phi(R,\mathbf{r})} S^{|R|\pm}_{R^{-1}\mathbf{r}} \nonumber\\
\mathcal{T}: & S^z_{\mathbf{r}} \rightarrow - S^z_{\mathbf{r}} \nonumber\\
& S^\pm_{\mathbf{r}} \rightarrow (-1)^n S^{\mp}_{\mathbf{r}} \nonumber\\
& i\to -i \nonumber\\
\mathcal{I}:&S^z_{\mathbf{r}}\to S^z_{\mathcal{I}^{-1}\mathbf{r}}\nonumber \\ 
&S^\pm_{\mathbf{r}}\to S^\pm_{\mathcal{I}^{-1}\mathbf{r}}
\label{eqs:doublet_spin_symmetry_transformations}
\end{align}
Here, $\mathbf{r}$ is the position of the pyrochlore lattice site. 
The phase $e^{\mp i \phi(R,\mathbf{r})}$ can be chosen to be one of the cube roots of unity with the right choice of basis. In the case of dipolar-octupolar doublets ($n=3$), the right choice of local basis gives $e^{\mp i n \phi(R,\mathbf{r})}=1$.
Note that, for spins on the pyrochlore lattice, inversion leaves a pyrochlore lattice site invariant, while the rotational symmetries leave a diamond site (the center of a tetrahedron in pyrochlore) invariant. 

In \cref{sec:nearest_neighbor_hamiltonian}, we define modified mirror symmetries $\mathcal{\tilde{\mathcal{M}}}$:
\begin{align}
    \tilde{\mathcal{M}}_R: & S^z_r \rightarrow S^z_{R^{-1}r} \nonumber\\
    & S^\pm_r \rightarrow e^{\pm in \phi(R,r)} S^{\pm}_{R^{-1}r} 
    \label{eq:define_modified_mirror_symmetry}
\end{align}
where, $R$ is a improper rotation of the $T_d$ point group. Note that the phase in the above transformation is the conjugate of that acquired during the the action of $R$ on the spins.

\section{Low-energy perturbative expansion}
\label{sec:low_energy_perturbative_expansion}

We consider the classical limit of the system with the Hamiltonian in \cref{eq:H_class3_2spin_A4} obtained by setting all parameters except $J_{zz}$ and $\kappa_{zz}$ to zero.
We define $V$ as the sum of all the remaining terms of the considered Hamiltonian:
\begin{align}
    H&=H_{zz} + V &\text{ with } &H_{zz} = \sum_{\sigma\in\{A,B\}}\sum_{\braket{ ij}_\sigma} J_{zz,\sigma} S_i^z S_j^z 
\end{align}
When $V=0$ and $\kappa_{zz}< 1$, the 2I2O ice states form the space of highly degenerate ground states. We define $P$ as a projector onto this spin-ice manifold and $Q=1-P$ as the projector to its orthogonal complement.
The perturbative expansion for the effective low-energy Hamiltonian of the system is:
\begin{align}
    H_{\text{eff}}= H_{zz} + P V P + P V G V P + P V G V G V P + ...
    \label{eq:perturbative_expansion}
\end{align}
where, 
\begin{align}
    G= \sum_{i \in Q} \frac{|i\rangle\langle i|}{H-{E_i}}
\end{align}
and $E_i$ is the energy of the state $|i\rangle$. We have used the notation $i\in Q$  to indicate that the state $|i\rangle$ is one of the orthonormal states in the space that $Q$ projects into.

%
%
\begin{figure}[b]
    \centering
    \includegraphics[width=.8\linewidth]{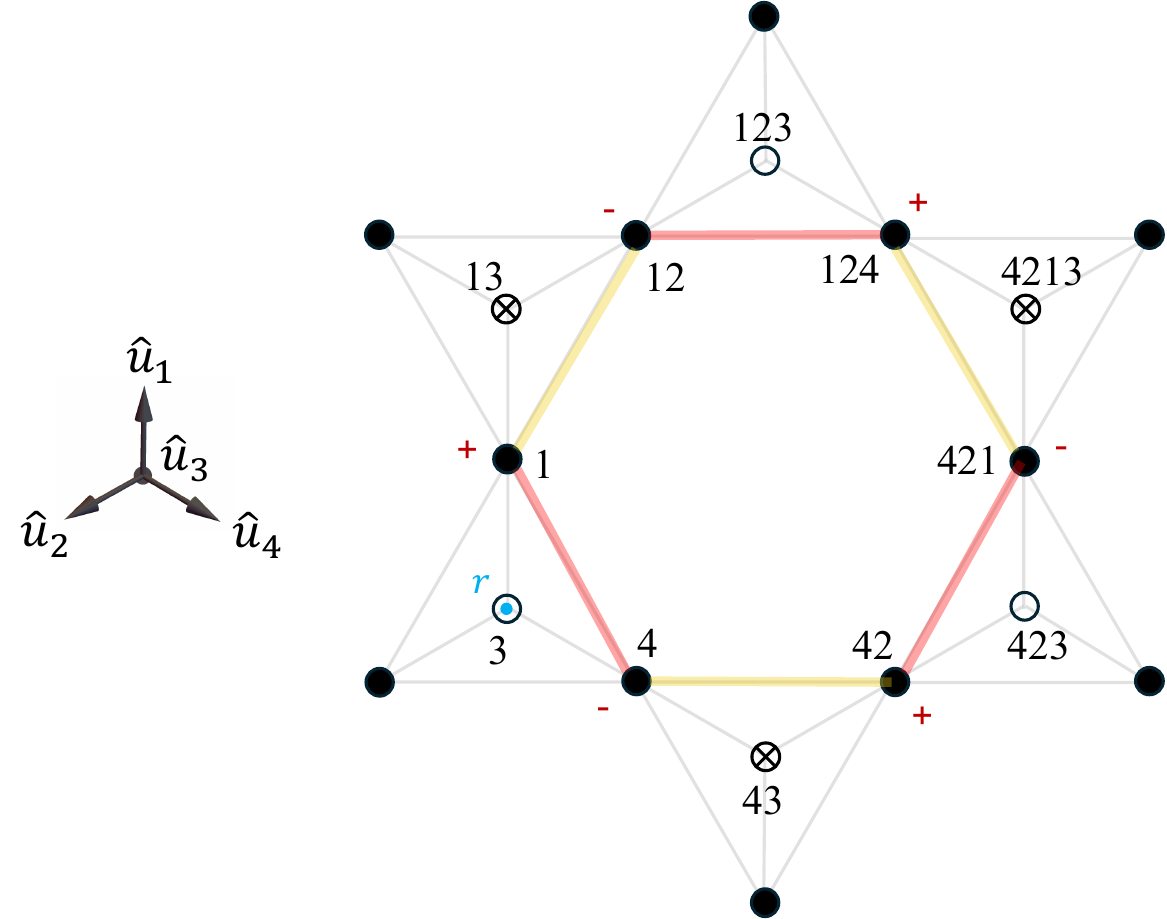}
    \caption{A section of the pyrochlore lattice showing a single hexagon plaquette. The vector $\hat{u}_3$ points out of the page. The sites drawn as an empty circle (crossed circle) indicate sites that are in the plane above (below) the plane of the page. The center of the bottom left upward-facing tetrahedron is chosen as the reference point $\mathbf{r}$, and the pyrochlore lattice sites are labeled by moving away from the reference point along the diamond lattice. The edges highlighted in red (yellow) show edges in sublattice A(B) along the hexagon. The $\pm$-signs at the lattice points along the hexagon align with the choice of $S^\pm$ at the sites in \cref{eq:eb_target_term}. (Color online.)  }
    \label{fig:pyro_hexagon_labeled}
\end{figure}
%
%

\section{NNN coupling generate $Y$-term} 
\label{sec:nnn_terms_and_generating_eb}

In this section, we calculate the perturbative expansion of the effective low energy Hamiltonian of the system with NNN coupling ($H=H_0+H_2+H_3$, see \cref{eq:H3_class3_A4,eq:H2_class3_NNN_A4,eq:H_class3_2spin_A4}) to compute the $Y$-coefficient of seven-spin terms of the form in \cref{eq:ham_yterm}.

In the main text, we use a compact indexing scheme, referring to spins either relative to the hexagonal plaquettes they lie on (see \cref{fig:pyrochlore_and_ice}) or by the NN pairs and NNN triplets they belong to (see \cref{fig:nnn_rl_parameter_description}).
This notation is hopefully familiar and intuitive to most readers. 
For the discussion in this appendix, it will prove more useful to refer to spins in relation to a reference point on the diamond lattice. As explained in \cref{fig:pyro_hexagon_labeled}, the spins are indexed by the directional hops required to reach them from the reference point.

\begin{widetext}

Consider one of the terms in the sum in \cref{eq:ham_yterm}, which involves point $\br$ in the $A$-sublattice of the diamond lattice (see \cref{fig:pyro_hexagon_labeled}):
\begin{align}
i Y(S^+_{1} S^-_{12} S^+_{124} S^-_{421} S^+_{42} S^-_{4}) \left(S^z_{3}+ S^z_{13}+S^z_{123} +S^z_{4213} +S^z_{423} +S^z_{43}\right)
\label{eq:eb_target_term}
\end{align}
Here, $S_\mu=S_{\mathbf{r},\mathbf{r}+ l\hat{u}_\mu}$ denotes the spin at the midpoint $\mathbf{r}+l\hat{\mu}/2$, $S_{\mu\nu}=S_{\mathbf{r}+l\hat{u}_\mu,\mathbf{r}+l\hat{u}_\mu-l\hat{u}_\nu}$ denotes the spin at the point  $\mathbf{r}+l\hat{\mu}-l\hat{\nu}/2$ and so on.

\subsection{Derivation of $Y_{\pm z}$ and $Y_{\pm +}$}
\label{sec:derivation_of_Ypmz_and_Ypmp}

To see that the $Y$-term is a part of the effective low energy description of the Hamiltonian $H=H_0+H_3$, we try to generate the term in \cref{eq:eb_target_term} perturbatively (see \cref{sec:low_energy_perturbative_expansion}). There are two different ways of perturbatively generating this term from this Hamiltonian, which we outline below.

Consider perturbatively generated terms of the form (see \cref{fig:pyro_hexagon_labeled})
\begin{align}
 \left( i\,t_{\mp z,l,B} \ S^-_{12} S^+_{1}S^z_{3} \right)\frac{1}{\left(J_{zz,A} \right)} \left( J_{\pm,B} \ S^-_{4} S^+_{42} \right)\frac{1}{\left(J_{zz,A} \right)} \left( J_{\pm,B} \ S^-_{421} S^+_{124} \right).
 \label{eq:example_tmpz_1}
\end{align}
These terms get added to the terms of the form
\begin{align}
 \left( i\,t_{\mp z,r,B} \ S^+_{42} S^-_{4}S^z_{3} \right) \frac{1}{\left(J_{zz,A} \right)}\left( J_{\pm,B} \ S^+_{1} S^-_{12} \right)\frac{1}{\left(J_{zz,A} \right)} \left( J_{\pm,B} \ S^+_{124} S^-_{421} \right).
\end{align}
Summing up all such terms on the considered hexagonal plaquette, i.e. seven-spin $Y$-terms containing $t_{\mp z}$, we get:
\begin{align}
&i 6 \frac{\left( t_{\mp z,r,B}+t_{\mp z,l,B} \right) \left(J_{\pm,B} \right)^2} {\left(J_{zz,A} \right)^2} 
(S^+_{1} S^-_{12} S^+_{124} S^-_{421} S^+_{42} S^-_{4}) (S^z_{3}+ S^z_{123} +S^z_{423}) \nonumber\\
&\ -i 6  \frac{\left( t_{\mp z,r,A}+t_{\mp z,l,A} \right) \left(J_{\pm,A} \right)^2} {\left(J_{zz,B} \right)^2}  
(S^+_{1} S^-_{12} S^+_{124} S^-_{421} S^+_{42} S^-_{4})  ( S^z_{13}  +S^z_{4213} +S^z_{43})
\end{align}
The factor of $6$ in front can be understood by noting that the operators in \cref{eq:example_tmpz_1} can be permuted to obtain other terms that have the same parameter coefficient. 
Extracting the coefficient of the term in \cref{eq:eb_target_term}, we get 
the expression for $Y_{\pm z}$ in \cref{eq:coeff_Y_pmz}.

Another way to perturbatively generate terms of the form in \cref{eq:eb_target_term} is to consider
\begin{align}
\frac{1}{\left(J_{zz}\right)^2} \left( -i\,t_{\mp -,l} \ S^+_{1} S^-_{12}S^+_{124} \right) \left( h_z S^z_{3} \right) \left( t_{\pm +,r} \ S^-_{4} S^+_{42}S^-_{421} \right)  ,
\end{align}
which, gets added to the terms of the form
\begin{align}
\frac{1}{\left(J_{zz}\right)^2} \left(-\ t_{\pm +,l} \ S^+_{1} S^-_{12}S^+_{124} \right) \left( h_z S^z_{3} \right) \left(-i\,t_{\mp-,r} \ S^-_{4} S^+_{42}S^-_{421} \right).
\end{align}
Adding up all such terms and extracting the coefficient in this case gives:
\begin{align}
    Y_{\pm +} &=  -\frac{6h_z \left( t_{\mp-,l}\  t_{\pm+,r} -t_{\mp-,r} \ t_{\pm+,l} \right) } {J_{zz}^2} \nonumber\\
\end{align}
The above expression and definitions in \cref{eq:define_bold_complex_parameters} lead to the expression in \cref{eq:coeff_Y_pmp}.

\subsection{Derivation of $Y_{2z+}$ and $Y_{2++}$}

We now consider the Hamiltonian $H=H_0+H_2$, and again look at the different ways to perturbatively generate the term in \cref{eq:eb_target_term}.

One way to generate the term is to consider:
\begin{align}
&\left(  J_{2z+,r,A} \ S^z_{3} S^-_{12} \right) \frac{1}{\left(J_{zz}\right)}  \left( J_{\pm,B} \ S^-_{421} S^+_{124} \right)\frac{1}{\left(J_{zz}\right)} \left( J_{\pm,B} \ S^-_{4} S^+_{42} \right)\frac{1}{\left(J_{zz}\right)} \left( i \frac{h_y}{2} S^+_{1} \right).
\end{align}
When adding up all terms of the above form, i.e. containing $J_{2z+,r,A}$ and $h_y$, unlike the cases in \cref{sec:derivation_of_Ypmz_and_Ypmp}, the terms with the same spin operators involved can have different denominators, depending on the order in which the operators are considered. This is due to the different energies of the intermediate states of the perturbative calculation. Also, changing the position where the operator $h_y\, S^+$ acts 
can change whether a $J_{\pm,B}$ or $J_{\pm,A}$ is required. Examples of each of these cases are shown below:
\begin{align}
&  \left(  J_{2z+,r,A} \ S^z_{3} S^-_{12} \right) \frac{1}{\left(J_{zz}\right)} \left( i \frac{h_y}{2} S^+_{1} \right)\frac{1}{\left(J_{zz,A}\right)}  \left( J_{\pm,B} \ S^-_{421} S^+_{124} \right)\frac{1}{\left(J_{zz,A}\right)} \left( J_{\pm,B} \ S^-_{4} S^+_{42} \right) \nonumber\\
&\ +\left(  J_{2z+,r,A} \ S^z_{3} S^-_{12} \right) \frac{1}{\left(J_{zz}\right)}  \left( J_{\pm,B} \ S^-_{421} S^+_{124} \right)\frac{1}{\left(J_{zz}\right)} \left( J_{\pm,A} \ S^-_{4} S^+_{1} \right)\frac{1}{\left(J_{zz}\right)} \left( i \frac{h_y}{2} S^+_{42} \right) 
\end{align}
There are 72 terms containing $J_{2z+,r,A}$ and $h_y$ ($4!$ ways to rearrange the operators for each one of three choices for the position of $h_y\, S^+$). 
Adding these up, we get
\begin{align}
    C(\kappa_{\pm},\kappa_{zz})  \frac{i J_{2z+,r,A}\, h_y\, J_{\pm}^2 \ }{2\, J_{zz}^3} (S^+_{1} S^-_{12} S^+_{124} S^-_{421} S^+_{42} S^-_{4}) \ S^z_{13}
\end{align}
where, $C(\kappa_{\pm},\kappa_{zz})$ is given by \cref{eq:define_C_kpm_kzz}. In the above analysis, note that we have disregarded terms where the two NN $J_\pm \, S^+S^-$ operators are replaced by NNN operators $J_{2++} \, S^+S^+$ and $J_{2++} \, S^-S^-$, or operators $iJ_{2--} \, S^+S^+$ and $iJ_{2--} \, S^-S^-$. We do this to simplify the calculation and note that we would expect $J_\pm$ to be much greater than $J_{2++}$ and $J_{2--}$ in a system with the $\mathcal{I}$, $\mathcal{T}$ and  $\mathcal{M}$ symmetries broken weakly.

Other ways of generating the $Y$-term using the NNN $S^z S^+$ and $S^z S^-$ operators are summarized by the the three terms below:
\begin{align}
&+\left(  J_{2z+,l,A} \ S^z_{3} S^+_{42} \right) \frac{1}{\left(J_{zz}\right)}  \left( J_{\pm,B} \ S^-_{421} S^+_{124} \right)\frac{1}{\left(J_{zz}\right)} \left( J_{\pm,B} \ S^-_{1} S^+_{12} \right)\frac{1}{\left(J_{zz}\right)} \left( -i \frac{h_y}{2} S^-_{4} \right) \nonumber\\
&\ +\left( -i\,J_{2z-,r,A} \ S^z_{3} S^-_{12} \right) \frac{1}{\left(J_{zz}\right)}  \left( J_{\pm,B} \ S^-_{421} S^+_{124} \right)\frac{1}{\left(J_{zz}\right)} \left( J_{\pm,B} \ S^-_{4} S^+_{42} \right)\frac{1}{\left(J_{zz}\right)} \left( \frac{h_x}{2} S^+_{1} \right) \nonumber \\  
&\ +\left( i\,J_{2z-,l,A} \ S^z_{3} S^+_{42} \right) \frac{1}{\left(J_{zz}\right)}  \left( J_{\pm,B} \ S^-_{421} S^+_{124} \right)\frac{1}{\left(J_{zz}\right)} \left( J_{\pm,B} \ S^-_{1} S^+_{12} \right)\frac{1}{\left(J_{zz}\right)} \left( \frac{h_x}{2} S^-_{4} \right)
\end{align}
Adding up all such terms, we obtain:
\begin{align}
&i \left( C(\kappa_{\pm},\kappa_{zz})\frac{  J_{\pm}^2}{2J_{zz}^3}   \left(h_y\, (J_{2z+,r,A} -J_{2z+,l,A})- h_x\,(J_{2z-,r,A} -J_{2z-,l,A}) \right)  \right) (S^+_{1} S^-_{12} S^+_{124} S^-_{421} S^+_{42} S^-_{4}) (S^z_{3}+ S^z_{123} +S^z_{423}) \nonumber\\
&\ -i \left(  C(\kappa_{\pm},\kappa_{zz})\frac{J_{\pm}^2}{2J_{zz}^3}   \left(h_y\, (J_{2z+,r,B} -J_{2z+,l,B})- h_x\,(J_{2z-,r,B} -J_{2z-,l,B}) \right)  \right) (S^+_{1} S^-_{12} S^+_{124} S^-_{421} S^+_{42} S^-_{4}) (S^z_{13}+ S^z_{4123} +S^z_{43})
\end{align}
Extracting the coefficient of the term in \cref{eq:eb_target_term} from the above expression gives \cref{eq:coeff_Y_2zp}. 

Lastly, we consider terms of the form 
\begin{align}
 \left( J_{2++,r}\ S_1^+ S_{42}^+ \right)\frac{1}{\left(2J_{zz}\right)}\left( h_z S^z_{3} \right)\frac{1}{\left(2J_{zz}\right)}\left( -iJ_{2--,l}\ S_{14}^- S_{421}^- \right)\frac{1}{\left(J_{zz,B}\right)} \left( J_{\pm,A}\ S_{12}^- S_{124}^+ \right)
\end{align} 
and
\begin{align}
\left(i J_{2--,r}\ S_1^+ S_{42}^+ \right)\frac{1}{\left(2J_{zz}\right)} \left( h_z S^z_{3} \right)\frac{1}{\left(2J_{zz}\right)} \left( J_{2++,l}\ S_{14}^- S_{421}^- \right)\frac{1}{\left(J_{zz,B}\right)} \left( J_{\pm,A}\ S_{12}^- S_{124}^+ \right).
\end{align}
When considering all terms with the above form, we again need to keep track of the energies of the intermediate states and account for the different positions that operators can act at. Taking the appropriate sum of of the 72 terms gives:
\begin{align}
    -i  \left(\left( 18\frac{(1+\kappa_{\pm}\kappa_{zz})}{1-\kappa_{zz}^2}+3 \right)  \frac{\left( J_{2--,l}\  J_{2++,r} -J_{2--,r} \ J_{2++,l} \right) J_{\pm} h_z } {\left(J_{zz}\right)^3} \right) &(S^+_{1} S^-_{12} S^+_{124} S^-_{421} S^+_{42} S^-_{4}) \nonumber\\
    &\ (S^z_{3}+ S^z_{13} +S^z_{123} +S^z_{4213} +S^z_{423} +S^z_{43})
\end{align}
The above expression gives us the coefficient $Y_{2++}$ in \cref{eq:coeff_Y_2pp}

\end{widetext}

\section{Magnetoelectric response of QSI}
\label{app:magentoelectric_response_of_QSI}

In this section, we show that an emergent ${\theta'}$ coupling of the $\mathbf{e}$ and $\mathbf{b}$ fields in QSI leads to an effective $\theta$ coupling of native $\mathbf{E}$ and $\mathbf{B}$ fields. In other words, a piece of QSI with an emergent $\theta'$ coupling is magnetoelectric. To prove this, we first study the steady-state solutions of a generic magnetoelectric sphere and then show that it matches the steady-state solution of a sphere of QSI with a $\theta'$ coupling only between the emergent fields.

\subsection{Coupling the emergent and native fields}

The spins in QSI are coupled to external native electromagnetic fields in the same way as one might expect for a generic insulator. An external magnetic field magnetizes a QSI sample, while an external electric field polarizes the sample. This interaction between the fields can be translated to the cross terms in the effective action of the emergent electromagnetic fields in QSI and the native electromagnetic fields, resulting in a joint action of the form:
\begin{align}
\nonumber 
 S  &= \int dt\ d^3 x \left( \frac{1}{8\pi \alpha c} \left( \mathbf{E}^2-c^2 \mathbf{B}^2 \right)
+\frac{1}{8\pi \alpha' {c'}} \left(\mathbf{e}^2 -{c'}^2 \mathbf{b}^2 \right) \right. \\
&- \left. \frac{\theta'}{4\pi^2} \mathbf{e}\cdot \mathbf{b}  
- g_{1} \ \mathbf{e}\cdot \mathbf{B}- g_{2} \ \mathbf{b}\cdot \mathbf{E}\right)
\label{eq:joint_action}
\end{align}
This expression contains the expected dominant terms of the action in the presence of all the usual symmetries \cite{laumannHybridDyonsInverted2023}. $g_1$ and $g_2$ are dimensionless parameters for the couplings between emergent and native fields. Note that we have only included a $\theta'$ coupling between the emergent fields in this expression and that the emergent fine structure constant, speed of light, and $\theta$ are denoted by $\alpha'$, ${c'}$, and $\theta'$ -- reserving $\alpha$, $c$, and $\theta$ for parameters associated with the native fields. 

The native and emergent fields can be expressed in terms of scalar and vector potentials as
\begin{align}
    \mathbf{E}&=- \boldsymbol{\nabla}\phi -\dot{\mathbf{A}}, &\mathbf{B}&=\boldsymbol{\nabla} \times \mathbf{A}.
    \label{eq:EB_potentials}\\
    \mathbf{e}&=-\boldsymbol{\nabla}\varphi'-\dot{\mathbf{a}} \quad \text{and} \quad &\mathbf{b}&=\boldsymbol{\nabla}\times \mathbf{a}.
    \label{eq:eb_potentials}
\end{align}
With these potentials, the action in \cref{eq:joint_action} can be used to obtain the following Euler-Lagrange equations of motion:
\begin{widetext}
\begin{equation}
\begin{aligned}
\frac{1}{4\pi\alpha c}\boldsymbol{\nabla}\cdot\mathbf{E} &= \boldsymbol{\nabla}\cdot(g_{2}\mathbf{b}) \\
\frac{1}{4\pi\alpha } \left(-\frac{1}{c} \dot{\mathbf{E}} + c\ \boldsymbol{\nabla} \times \mathbf{B} \right)&=- \partial_t(g_{2}\mathbf{b})-\boldsymbol{\nabla}\times(g_{1}\mathbf{e})\\
\frac{1}{4\pi}\boldsymbol{\nabla}\cdot \left( \frac{\mathbf{e}}{\alpha' {c'}}-\frac{\theta'\mathbf{b}}{\pi} \right) &=  \boldsymbol{\nabla}\cdot(g_{1}\mathbf{B}) \\
\frac{1}{4\pi } \left(  -\partial_t\left( \frac{\mathbf{e}}{{\alpha'c'}} -\frac{\theta'\mathbf{b}}{\pi} \right) + \boldsymbol{\nabla} \times\left(  \frac{c'\mathbf{b}}{\alpha'} +\frac{\theta'\mathbf{e}}{\pi}\right) \right) &= - \partial_t (g_{1}\mathbf{B})-\boldsymbol{\nabla}\times(g_{2} \mathbf{E})
\end{aligned}
\label{eq:coupled_EB_eb_eoms}
\end{equation}
\end{widetext}
The fields expressed in terms of the potentials as in \cref{eq:EB_potentials,eq:eb_potentials} also lead to the homogeneous Maxwell equations for the native and emergent fields:
\begin{align}
\boldsymbol{\nabla}\cdot\mathbf{B}&=0 \quad\text{and} \quad &
\dot{\mathbf{B}} + \boldsymbol{\nabla}\times \mathbf{E}&=0.
\label{eq:EB_bianchi}\\
\boldsymbol{\nabla}\cdot\mathbf{b}&=0 \quad \text{and} \quad &\dot{\mathbf{b}}+\boldsymbol{\nabla}\times \mathbf{e}&=0
\label{eq:eb_bianchi}
\end{align}

While the divergence-free condition on the emergent magnetic field is not a necessary imposition, we limit ourselves to the case with no emergent monopoles as we are interested in the low-energy physics of the problem. 

Notice that the equations in \cref{eq:coupled_EB_eb_eoms} contain boundary terms -- terms that only appear where the spatial derivative of $\theta'$, $g_1$, or $g_2$ is non-zero. Spatial boundaries of a QSI material can be modeled by a decaying spatial variation of the parameters $\theta,g_1$ and $g_2$. (We will only consider parameters that are static here and, therefore, assume that all time derivatives of parameters are zero.)

We consider a step function variation of these parameters along the boundary of a sphere and show (in \cref{sec:emergent_fields_in_a_sphere_coupled_to_native_fields}) that QSI with a non-zero emergent $\theta'$ exibits a magnetoelectric response to external native electromagnetic fields. Applying a uniform external magnetic field leads to polarization of the QSI sample and a dipolar electric field outside the sphere. Similarly, an external electric field leads to the magnetization of the QSI sample and a dipolar magnetic field outside the sphere. When the dimensionless parameters $\theta'$ is small, the response of the sphere can be summarized by assigning it an effective $\theta$-term coupling of the native fields with 
\begin{align}
\theta=-\frac{16\pi^2{\alpha'}^2}{\left( 1+32\pi^2\, \frac{v}{c} \alpha\alpha' g_{1}^2 \right) \left( 1+16 \pi^2\, \frac{c}{v} \alpha\alpha'g_{2}^2 \right)  }\,g_{1}g_{2}\,\theta'
\label{eq:effective_theta}
\end{align}
This shows that an emergent $\theta$-term does lead to a magnetoelectric response of QSI if \emph{both} $g_1$ and $g_2$ are non-zero. 

Since ${\cal T}$ and ${\cal I}$ are broken to generate $\theta'$, one could also include coupling of the form $\mathbf{e}\cdot \mathbf{E}$ and $\mathbf{b}\cdot \mathbf{B}$ in the Lagrangian in \cref{eq:joint_action}. 
The presence of a magnetoelectric response does not necessarily indicate a $\theta'$-coupling between the emergent fields, as it could be generated by these other couplings.
However, in this article, we present the simplest way an emergent $\theta'$ can lead to an effective $\theta$, and therefore, we constrain our analysis to couplings between native and emergent fields that are ${\cal T}$ and ${\cal I}$ symmetric.

\subsection{A magnetoelectric sphere}

Before we analyze the full set of coupled equations for QSI (\cref{eq:coupled_EB_eb_eoms}), we shall consider the system of a magnetoelectric sphere in a vacuum governed by the action 
\begin{equation}
    S  = \int dt\ d^3 x \left( \frac{1}{8\pi \alpha c} \left( \mathbf{E}^2-c^2 \mathbf{B}^2 \right)
    - \frac{\theta}{4\pi^2} \mathbf{E}\cdot \mathbf{B} \right)
    \, .
\end{equation}
This is simply the action of the native electromagnetic fields in the presence of an arbitrary magnetoelectric material parametrized by $\theta$, which is uniform and non-zero inside the sphere and zero outside the sphere. 
The Euler-Lagrange equations of motion (EOMs) of the system are:
\begin{align}
    \frac{1}{c}\boldsymbol{\nabla} \cdot \mathbf{E} &=  \frac{\alpha}{\pi}\boldsymbol{\nabla}\theta \cdot \mathbf{B}\\
    -\frac{1}{c^2} \dot{\mathbf{E}} +\boldsymbol{\nabla} \times \mathbf{B} &= - \frac{\alpha}{\pi c} \left(\dot\theta \ \mathbf{B} + \boldsymbol{\nabla}\theta \times \mathbf{E} \right) 
    \, .
\end{align}
Notice that steady-state solutions in the bulk inside the sphere and outside the sphere are unaffected by $\theta$. However, the boundary conditions for the fields are modified due to the spatial step function variation of $\theta$.
The fields also satisfy the homogeneous Maxwell equations (as a consequence of the Bianchi-identities) in \cref{eq:EB_bianchi}.

The EOMs can be simplified by defining the auxiliary magnetic field, $\mathbf{H}$, and electric displacement field, $\mathbf{D}$, as
\begin{equation}
    \mathbf{H}=\frac{c}{4\pi\alpha} \mathbf{B} + \frac{\theta}{4\pi^2} \mathbf{E} \quad \text{and } \quad\mathbf{D}= \frac{1}{4\pi\alpha c}\mathbf{E} - \frac{\theta}{4\pi^2} \mathbf{B}
    \, .
\end{equation}
The EOMs, in terms of these fields, are
\begin{equation}
    \boldsymbol{\nabla}\cdot \mathbf{D}=0 \quad\text{and} \quad -\dot{\mathbf{H}}+\boldsymbol{\nabla}\times \mathbf{D} =0 
    \, .
\end{equation}
To find the steady-state solutions, we work with the fields that have zero curl everywhere (including the boundary):  $\mathbf{E}$ and $\mathbf{H}$.
These fields have boundary conditions:
\begin{equation}
\begin{aligned}
\mathbf{E}^{\parallel}_{\text{in}} &=\mathbf{E}^{\parallel}_{\text{out}}\\ \mathbf{H}^\parallel_{\text{in}} &=\mathbf{H}^\parallel_{\text{out}} \\
E^\perp_{\text{in}} -E^{\perp}_{\text{out}} &=\frac{\alpha c}{\pi}\theta\ B^\perp= 4\alpha^2\theta\ H^\perp _{\text{out}} \\
H^\perp_{\text{in}} -H^{\perp}_{\text{out}} &=\frac{1}{4\pi^2} \theta \ E^\perp _{\text{in}} 
\end{aligned}
\label{eq:sphere_EM_BC}
\end{equation}
where, $\mathbf{E}^\parallel$ is the field parallel to the boundary and $E^\perp$ is the component perpendicular to the boundary.
We get the first two conditions from zero curls of $\mathbf{E}$ and $\mathbf{H}$, and the next two from zero divergences of $\mathbf{D}$ and $\mathbf{B}$ at the boundary.

Since these fields have zero curl, we can define them as gradients of scalar fields:
\begin{align}
\mathbf{E}&=-\boldsymbol{\nabla}\phi, &\mathbf{H}&=-\boldsymbol{\nabla}W
\label{eq:EH_in_terms_PhiW}
\end{align}
Everywhere but the boundary, these fields have zero divergences. Hence, the scalar fields in the bulk will be solutions of the Laplace equation.

We are interested in solutions that give uniform electric ($E_0$) and magnetic ($H_0$) fields far from the sphere. For such solutions, we have:
\begin{align}
    \phi_{\text{in}} &= -E_{\text{in}} \ r \cos\vartheta,& \phi _{\text{out}} = \left( -E_0\ r +{E_{\text{d}}}\frac{ R^3}{r^2} \right)\cos\vartheta,\nonumber\\
    W_{\text{in}} &=-H_{\text{in}} \ r \cos\vartheta,&W _{\text{out}} = \left( -H_0\ r +{H_{\text{d}}}\frac{R^3}{r^2} \right)\cos\vartheta
    \, .
\label{eq:PhiW_generic_Laplace_solutions}
\end{align}
Here, $R$ is the radius of the sphere, $r$ is the radial distance and $\vartheta$ is the polar angle in spherical coordinates with the center of the sphere chosen to be the origin.
These give us uniform fields in the sphere and fields with a dipolar component outside the sphere:
\begin{align}
    \mathbf{E}_{\text{in}} &= E_{\text{in}} \ \hat{z},&
    \mathbf{E}_{\text{out}} &= E_{0}\ \hat{z} +{E_{\text{d}}} \frac{R^3}{r^3} \left( 2\cos \vartheta\  \hat{r}- \sin\vartheta\ \hat{\vartheta} \right) \nonumber \\ 
    \mathbf{H}_{\text{in}} &= H_{\text{in}} \ \hat{z},&
    \mathbf{H}_{\text{out}} &= H_{0}\ \hat{z} + {H_{\text{d}}}\frac{R^3}{r^3} \left( 2\cos \vartheta\  \hat{r}- \sin\vartheta\ \hat{\vartheta} \right) 
\label{eq:EH_generic_solutions}
\end{align}
The four constants $E_{\text{in}},\,E_{\text{d}},\,H_{\text{in}}$ and $H_{\text{d}}$ can be determined by using the boundary conditions specified in \cref{eq:sphere_EM_BC}. These give the relations:
\begin{equation}
    \begin{aligned}
        E_{\text{in}} &=E_{0}-{E_{\text{d}}} \\
        H_{\text{in}} &=H_0-{H_{\text{d}}}\\
        E_{\text{in}}-\left( E_{0}+2{E_{\text{d}}} \right) &=4\alpha^2\theta \left( H_{0}+2{H_{\text{d}}} \right) \\
        H_{\text{in}}-\left( H_{0}+2{H_{\text{d}}}\right)   &= \frac{\theta}{4\pi^2} E_{\text{in}} 
    \end{aligned}
\end{equation}
These can be solved to obtain:
\begin{equation}
\begin{aligned}
    {E}_{\text{in}}&= \frac{1 }{\left( 1+ \frac{2\alpha ^2 }{9\pi^2}\theta ^2 \right)} \left( {E}_0+\frac{4}{3} \alpha ^2 \theta {H_{0}}\right)\\
    {E_{\text{d}}}&=\frac{-1 }{\left( 1+ \frac{2\alpha ^2 }{9\pi^2}\theta ^2 \right)}   \left(\frac{4 \alpha ^2 }{3} \theta  {H_{0}}-\frac{2 \alpha ^2 }{9 \pi^2} \theta^2{E}_0  \right)\\
    H_{\text{in}}&= \frac{1 }{\left( 1+ \frac{2\alpha ^2 }{9\pi^2}\theta ^2 \right)}  \left( {H_{0}} \left(1+\frac{\alpha ^2 \theta^2}{3\pi^2}\right)+\frac{1}{12\pi^2} \theta {E}_0  \right)\\
    {H_{\text{d}}}&= \frac{-1 }{\left( 1+ \frac{2\alpha ^2 }{9\pi^2}\theta ^2 \right)}    \left(\frac{1}{12\pi^2} \theta
    {E}_0+ \frac{\alpha ^2}{9\pi^2} \theta^2  {H_{0}}\right)
\end{aligned}
\end{equation}
The magnetic field is given by
\begin{align}
    \mathbf{B}_{\text{in}} &=  \frac{4\pi\alpha}{c}\mathbf{H}_{\text{in}} -\frac{\alpha\theta}{\pi c}\mathbf{E}_{\text{in}} =B_{\text{in}} \ \hat{z} 
    \label{eq:sphere_EB_Bin}\\ 
    \mathbf{B}_{\text{out}} &= \frac{4\pi\alpha}{c}\mathbf{H}_{\text{out}} =B_{0}\ \hat{z} + {B_{\text{d}}}\frac{R^3}{r^3} \left( 2\cos \vartheta\  \hat{r}- \sin\vartheta\ \hat{\vartheta} \right)
    \label{eq:sphere_EB_Bout}
\end{align}
where, $B_{0}=\frac{4\pi\alpha}{c}H_{0}$, ${B_{\text{d}}}=\frac{4\pi\alpha}{c}H_{\text{d}}$ and
\begin{align}
    B_{\text{in}} &=\frac{1 }{\left( 1+ \frac{2\alpha ^2 }{9\pi^2}\theta ^2 \right)} \left( {B_{0}} -\frac{ 2\alpha }{3\pi c}\theta  E_0   \right).
\end{align}
For small $\theta$, we arrive at:
\begin{equation}
\begin{aligned}
    E_{\text{in}} &=E_{0}+ \frac{\alpha c}{3\pi} \theta\, B_{0}+O(\theta^2)\\
    {E_{\text{d}}}&=-\frac{\alpha c}{3\pi} \theta\, B_{0}+O(\theta^2)\\
    B_{\text{in}} &=B_{0}- \frac{2\alpha }{3\pi c} \theta\, E_{0}+O(\theta^2)\\
    {B_{\text{d}}} &=- \frac{\alpha }{3\pi c} \theta\, E_{0}+O(\theta^2)
\end{aligned}
\label{eq:sphere_EM_solutions_expansion}
\end{equation}
The above equations show us that the sphere gains polarization in response to an external magnetic field and gains magnetization in response to an external electric field, as should be expected of a magnetoelectric sphere.
The uniform polarization and magnetization fields inside the sphere are accompanied by dipolar electric and dipolar magnetic fields, respectively, outside the sphere.

\subsection{Emergent fields in a sphere coupled to native fields }
\label{sec:emergent_fields_in_a_sphere_coupled_to_native_fields}

Let us now consider a sphere with emergent compact $U(1)$ gauge fields coupled to external fields. 
The emergent fields have a $\theta$-term, and the complete action of the system is given by \cref{eq:joint_action}. The constants $\theta',g_1$ and $g_2$ are defined and considered uniformly non-zero only inside the sphere. The Euler-Lagrange EOMs, in this case, are shown in \cref{eq:coupled_EB_eb_eoms}.
The homogeneous Maxwell equations in \cref{eq:EB_bianchi} are still valid everywhere and the equations in \cref{eq:eb_bianchi} are valid inside the sphere. 
Similarly to the last section, the EOMs can be simplified by defining the fields: 
\begin{equation}
\begin{aligned}
    \mathbf{H}&=\frac{c}{4\pi\alpha} \mathbf{B}+g_{1}\mathbf{e} \, , \\
    \mathbf{D}&=\frac{1}{4\pi{\alpha}} \frac{\mathbf{E}}{c}-g_{2}\mathbf{b} \, , \nonumber \\ 
    \mathbf{h}&=\frac{{c'}}{4\pi{\alpha'}} \mathbf{b}+\frac{\theta'}{4\pi^2} \mathbf{e}+g_{2}\mathbf{E} \, , \\
    \mathbf{d}&=\frac{1}{4\pi\alpha_{a}} \frac{\mathbf{e}}{{c'}}-\frac{\theta_{a}}{4\pi^2} \mathbf{b}-g_{1}\mathbf{B}
    \, .
\end{aligned}
\end{equation}
In terms of these fields, the EOMs in \cref{eq:coupled_EB_eb_eoms} are simply given by:
\begin{align}
    \boldsymbol{\nabla} \cdot \mathbf{D}&=0 \, ,
    &-\dot{\mathbf{D}}+\boldsymbol{\nabla}\times \mathbf{H}&=0 \, , \nonumber \\ 
    \boldsymbol{\nabla} \cdot \mathbf{d}&=0 \, , \quad\text{and} &
    -\dot{\mathbf{d}}+\boldsymbol{\nabla}\times \mathbf{h}&=0
    \, .
\end{align}

To find the steady-state solutions in this case, we work with fields $\mathbf{E}$ and $\mathbf{H}$, and the emergent fields with zero curl, $\mathbf{e}$ and $\mathbf{h}$.
The boundary conditions are again given by zero curls of fields $\mathbf{E}$ and $\mathbf{H}$, and zero divergences of $\mathbf{D}$ and $\mathbf{B}$:
\begin{equation}
\begin{aligned}
    \mathbf{E}^{\parallel}_{\text{in}} &=\mathbf{E}^{\parallel}_{\text{out}} \\
    \mathbf{H}^\parallel_{\text{in}} &=\mathbf{H}^\parallel_{\text{out}} \\
    E^\perp_{\text{in}} -E^{\perp}_{\text{out}} &=4\pi\alpha cg_{2}\ b^\perp\\
    &= 16\pi^2\alpha{\alpha'}\frac{c}{{c'}} g_{2}\left( h^\perp-\frac{\theta'}{4\pi^2} e^\perp-g_{2}\ E^\perp _{\text{in}}  \right)  \\
    H^\perp_{\text{in}} -H^{\perp}_{\text{out}} &=g_{1}e^\perp
\end{aligned}
\end{equation}
We also get boundary conditions from requiring zero curl of $\mathbf{h}$ and zero divergence $\mathbf{d}$:
\begin{equation}
\begin{aligned}
    \mathbf{h^\parallel}&=0\\
    \frac{1}{4\pi{\alpha'} {c'}}e^\perp &= \frac{\theta'}{4\pi^2} b^{\perp} + g_{1} B^\perp\\
    &= \frac{{\alpha'}\theta'}{\pi {c'}} \left(h^\perp -\frac{\theta'}{4\pi^2} e^\perp-g_{2}\ E^\perp _{\text{in}}   \right)  + g_{1} \frac{4\pi{\alpha'}}{c} H^\perp_{\text{out}}
\end{aligned}    
\end{equation}
Note that for emergent fields, we do not require $\boldsymbol{\nabla}.\mathbf{b}=0$ or $\boldsymbol{\nabla}\times \mathbf{e}=0$ at the boundary as these fields are only defined inside and on the surface of the sphere. 

The setup to find the steady-state solutions of $\mathbf{E}$ and $\mathbf{H}$ is the same as in \cref{eq:EH_in_terms_PhiW,eq:PhiW_generic_Laplace_solutions,eq:EH_generic_solutions}. To apply the same procedure for the fields $\mathbf{e}$ and $\mathbf{h}$, we use the fact that these fields have zero curl inside to define them in terms of scalar fields $\varphi$ and $w$ by:
\begin{align}
    \mathbf{e}&=-\boldsymbol{\nabla}\varphi \, , &\mathbf{h}&=-\boldsymbol{\nabla} w
\end{align}
Everywhere but the boundary, these fields have zero divergences and will be solutions of the Laplace equation, which only allows solutions of the form:
\begin{equation}
    \varphi = -e\ r \cos\vartheta,\quad w=-h\ r \cos\vartheta
\end{equation}
These solutions correspond to uniform fields inside the sphere:
\begin{align}
    \mathbf{e}&=e\ \hat{z} \, , \quad &\mathbf{h}&=h\ \hat{z}
    \, .
\end{align}

The boundary conditions give the relations:
\begin{equation}
\begin{aligned}
    E_{\text{in}} &=E_{0}-{E_{\text{d}}} \\
    H_{\text{in}} &=H_0-{H_{\text{d}}}\\
    E_{\text{in}}-\left( E_{0}+2{E_{\text{d}}} \right) &=16\pi^2\alpha{\alpha'} \frac{c}{{c'}}g_{2}\left( h-\frac{\theta'}{4\pi^2} e-g_{2}\ E_{\text{in}}  \right)  \\
    H_{\text{in}}-\left( H_{0}+2{H_{\text{d}}} \right)  &= g_{1}\, e\\
    h&=0\\
    \frac{1}{4\pi{\alpha'} {c'}} e&=\frac{{\alpha'}\theta'}{\pi {c'}} \left(  h-\frac{\theta'}{4\pi^2} e-g_{2}\ E_{\text{in}}  \right) \\&\quad + \frac{4\pi\alpha}{c} g_{1} \left( H_{0}+2H_d\right)  
\end{aligned}
\end{equation}

We first define some constants to make the solutions of the above equations easy to represent:
\begin{align}
    \tilde{\alpha}&= 16\pi^2\alpha{\alpha'} \\
    u&=\frac{{c'}}{c} \\
    \chi&=\left( 1+\frac{2}{3} u \tilde{\alpha } g_1^2 +\frac{\tilde{\alpha }}{3 u}g_2^2 +\frac{2}{9}  \tilde{\alpha }^2 g_1^2 g_2^2+\frac{{\alpha'}^2 }{\pi ^2} {\theta'}^2\right)^{-1}
\end{align}
The solutions are:
\begin{equation}
\begin{aligned}
    E_{\text{in}}&= \chi \left(1+\frac{2}{3} \tilde{\alpha }u g_1^2 +\frac{ {\alpha'}^2 }{\pi ^2} {\theta'}^2 \right) E_{0} -\chi\frac{ \tilde{\alpha }^2 }{12 \pi^2}g_1 g_2 \theta' H_0\\ 
    E_d&= \chi\left(1+\frac{2}{3} \tilde{\alpha}u g_1^2 \right) \frac{ \tilde{\alpha }}{3 u}g_2^2E_{0} +\chi\frac{\tilde{\alpha }^2 }{12 \pi ^2} g_1 g_2 \theta' H_0\\ 
    H_{\text{in}}&= \chi \left(1+ u \tilde{\alpha} g_1^2 +\frac{ \tilde{\alpha }}{3 u} g_2^2 +\frac{1}{3} \tilde{\alpha }^2 g_1^2 g_2^2   +\frac{ {\alpha'}^2 }{\pi^2} {\theta'}^2 \right) H_{0} \\&\quad -\chi\frac{4}{3}  {\alpha'}^2  g_1 g_2 \theta' E_0\\ 
    H_d&= -\chi \left(1 +\frac{\tilde{\alpha } }{3u} g_2^2  \right)\frac{1}{3}  \tilde{\alpha }u  g_1^2 H_{0}+\chi\frac{4}{3}   {\alpha'}^2   g_1 g_2\theta' E_0\\
    e&= \chi \left(1+\frac{\tilde{\alpha } }{3u} g_2^2 \right)\tilde{\alpha }u g_1 H_{0} -\chi4 {\alpha'}^2  g_2 \theta'E_0\\ 
    h&=0
\end{aligned}
\label{eq:sphere_em_solutions}
\end{equation}
The magnetic field inside the sphere, in this case, is given by
\begin{align}
    \mathbf{B}_{\text{in}} &= \frac{4\pi\alpha}{c}\mathbf{H}_{\text{in}}-\frac{4\pi\alpha}{c}g_{1}\mathbf{e}\, =B_{\text{in}} \ \hat{z} \, ,
\end{align}
with
\begin{align}
    B_{\text{in}} &= \chi \left(1+\frac{ \tilde{\alpha } }{3 u}g_2^2+\frac{ {\alpha'}^2}{\pi^2} {\theta'}^2 \right)B_{0}+\chi\frac{2  \tilde{\alpha } {\alpha'} }{3 \pi  c}  g_1 g_2 \theta'E_0
    \, .
\end{align}
The magnetic field outside is obtained by \cref{eq:sphere_EB_Bout}, with $B_{\text{d}}$ obtained from the solution for $H_{\text{d}}$ in \cref{eq:sphere_em_solutions}.
The emergent magnetic field is given by
\begin{align}
    \mathbf{b} &= \frac{4\pi\alpha'}{{c'}}\mathbf{h}-\frac{\alpha'\theta'}{\pi {c'}}\mathbf{e}-\frac{4\pi\alpha'}{{c'}}g_{2}\,\mathbf{E}_{\text{in}}= b \ \hat{z}  \, ,
\end{align}
where
\begin{align}
    b&= -\chi \left(1+\frac{2 }{3 } u \tilde{\alpha } g_1^2 \right) \frac{4 \pi {\alpha'}}{u} g_{2} E_{0} -\chi \,4  {\alpha'}^2 g_1 \theta' B_{0} \, .
\end{align}

The above expressions for the steady-state solutions are easier to probe when expanded under the condition $\theta'< 1$:
\begin{widetext}
\begin{align}
E_{\text{in}} &=\frac{3u}{\left( 3u+\tilde{\alpha} g_{2}^2\right)} E_{0} -\frac{48\pi c^2v \alpha\alpha'^2\,g_{1}g_{2} }{(3c+2v\tilde{\alpha}g_{1}^2)(3v+\tilde{\alpha}g_{2}^2)}\theta' B_{0} +O( {\theta'}^2 )\\
E_{\text{d}}&=\frac{\tilde{\alpha} g_{2}^2}{\left( 3u+\tilde{\alpha} g_{2}^2\right)} E_{0}+\frac{48\pi c^2v \alpha\alpha'^2\,g_{1}g_{2} }{(3c+2v\tilde{\alpha}g_{1}^2)(3v+\tilde{\alpha}g_{2}^2)}\theta' B_{0} +O({\theta'}^2)\\
B_{\text{in}} &=\frac{3}{\left( 3+2u\tilde{\alpha} g_{1}^2\right)} B_{0}+\frac{96\pi v \alpha\alpha'^2\,g_{1}g_{2} }{(3c+2v\tilde{\alpha}g_{1}^2)(3v+\tilde{\alpha}g_{2}^2)}\theta' E_{0} +O({\theta'}^2)\\
B_{\text{d}}&=-\frac{u\tilde{\alpha} g_{1}^2}{\left( 3+2u\tilde{\alpha} g_{1}^2\right)} B_{0}+\frac{48\pi v \alpha\alpha'^2\,g_{1}g_{2} }{(3c+2v\tilde{\alpha}g_{1}^2)(3v+\tilde{\alpha}g_{2}^2)}\theta' E_{0} +O({\theta'}^2)
\label{eq:sphere_em_solutions_expansion}
\end{align}
\end{widetext}
The above expressions demonstrate that the sphere behaves similarly to a magnetoelectric sphere. Comparing the above expression to those in \cref{eq:sphere_EM_solutions_expansion}, we can conclude that the sphere gains an effective $\theta$ from the coupling to emergent fields with an emergent $\theta'$.

When the dimensionless parameter $\theta'$ is small, the response of the sphere can be summarized by assigning it an effective $\theta$-term coupling of the native fields with 
\begin{align}
\theta=-\frac{16\pi^2{\alpha'}^2}{\left( 1+2 u \tilde{\alpha}g_{1}^2 \right) \left( 1+\frac{1}{u} \tilde{\alpha}g_{2}^2 \right)  }\,g_{1}g_{2}\,\theta'
\label{eq:effective_theta_compact}
\end{align}
This shows that an emergent $\theta$-term does lead to a magnetoelectric response of quantum spin ice if both $g_1$ and $g_2$ are non-zero. 



\input{citations.bbl}

\end{document}

%% file: citations.bbl
%